\documentclass[11pt,a4paper]{article} 
\usepackage{jheppub}
\usepackage[utf8]{inputenc}

\usepackage{graphicx}
\usepackage{amsmath}
\usepackage{amsfonts}
\newcommand{\beq}{\begin{equation}}
\newcommand{\eeq}{\end{equation}}
\newcommand{\beqa}{\begin{eqnarray}}
\newcommand{\eeqa}{\end{eqnarray}}
\def\lsim{\raise0.3ex\hbox{$<$\kern-0.75em\raise-1.1ex\hbox{$\sim$}}}
\def\gsim{\raise0.3ex\hbox{$>$\kern-0.75em\raise-1.1ex\hbox{$\sim$}}}

\def\x{{\bf x}}

\def\0{{\bf 0}}

\def\kk{{\kappa}}
\def\pp{{\hat{p}}}

\title{Rapidity dependence of heavy-flavour production in heavy-ion collisions within a full 3+1 transport approach: quenching, elliptic and directed flow}

\author[a]{Andrea Beraudo}
\author[a]{Arturo De Pace}
\author[a]{Marco Monteno}
\author[a]{Marzia Nardi}
\author[a]{Francesco Prino}

\affiliation[a]{INFN, Sezione di Torino, via Pietro Giuria 1, I-10125 Torino}

\emailAdd{beraudo@to.infn.it}
\emailAdd{depace@to.infn.it}
\emailAdd{monteno@to.infn.it}
\emailAdd{nardi@to.infn.it}
\emailAdd{prino@to.infn.it}

\abstract{We extend our POWLANG transport setup for the modelling of heavy-flavour production in heavy-ion collisions to the case of full 3+1 simulations, dropping the approximation of longitudinal boost-invariance of the background medium. This enables us to provide predictions for observables for which the rapidity dependence is essential in order to obtain a non-vanishing signal, like the directed flow $v_1$, and to get reliable results also for kinematic distributions of heavy-flavour particles at forward rapidity. We compare our predictions with experimental data obtained in Au-Au and Pb-Pb collisions at RHIC and at the LHC.}

\begin{document}

\maketitle

\section{Introduction}
Heavy-flavour (HF) hadrons and their decay products (electrons and muons) arising from parent charm and beauty quarks are among the best probes of the medium formed in relativistic heavy-ion collisions. Due to their large mass heavy quarks (HQ's) arise from hard scattering processes occurring very early after the collision, before the formation of a thermalized plasma of gluons and light quarks (QGP) and their initial production can be calculated using pQCD. Hence, before hadronizing they cross the fireball produced in the collision probing all the stages of its evolution, performing a tomography of the latter. Furthermore, theoretical estimates indicate that the relaxation time of HQ's is of the same order as the lifetime of the QGP, so that one expects that charm and beauty quarks approach only partial thermal equilibrium with the fireball through which they propagate: deviations from full thermalization have thus the potential to put phenomenological constraints on the values of the HQ transport coefficients.

Most HF measurements concern the nuclear modification factor $R_{\rm AA}$~\cite{Adam:2015sza,Adamczyk:2014uip} and the elliptic anisotropy $v_2$~\cite{Adare:2006nq,Acharya:2017qps}  of the momentum and angular distributions of charmed and beauty hadrons and their decay products. This allows one to get access to the quenching of the distributions at high momentum due to in-medium parton energy-loss and to the collective (radial and elliptic) flow acquired by the heavy quarks in the expanding fireball.
More recently HF studies have been extended to observables sensitive to event-by-event fluctuations of the initial geometry of the fireball, like the triangular flow $v_3$~\cite{Sirunyan:2017plt,Acharya:2020pnh,Aad:2020grf} and the event-shape-engineering analysis of the $D$-meson $v_2$~\cite{Acharya:2018bxo,Acharya:2020pnh}, both satisfactorly reproduced by several transport calculations~\cite{Nahrgang:2014vza,Prado:2016szr,Uphoff:2012gb,Song:2015ykw,He:2019vgs,Cao:2016gvr,Ke:2018tsh,Beraudo:2017gxw,Beraudo:2018tpr}.
Another interesting development concerns the modification of HF hadrochemistry in nucleus-nucleus collisions (and, more generally, in high-multiplicity hadronic collisions), with a relative enhancement of the production of $D_s$ and $B_s$ mesons and $\Lambda_c$ baryons relative to non-strange mesons observed by the ALICE~\cite{Acharya:2018hre,Acharya:2018ckj} and CMS~\cite{Sirunyan:2019fnc,Sirunyan:2018zys} collaborations at the LHC and by the STAR~\cite{Vanek:2019dqi,Adam:2019hpq,Adam:2021qty} collaboration at RHIC.

So far most theoretical calculations developed for the study of HF observables are based on a (2+1)D modelling of the background medium, assuming its invariance for longitudinal boosts: most experimental measurements are in fact performed around mid-rapidity and this justifies such an approximation which allows one to save computing and storage resources.
However, there are observables for which a full (3+1)D description of the medium is in order, either because measurements are performed at quite forward rapidity (e.g. the muons from HF decays measured by ALICE) or because their dependence on the particle rapidity is fundamental in order to get a non-vanishing signal (e.g. the directed flow $v_1$, which would be zero if integrated over a symmetric interval around mid-rapidity).
Hence in this paper we extend our POWLANG transport setup~\cite{Beraudo:2014boa} to correctly deal with the rapidity dependence of the observables. This is done interfacing the Langevin simulation of the HQ propagation to the full (3+1)D output of the ECHO-QGP code~\cite{DelZanna:2013eua} employed to model the evolution of the background medium, for which we take an initial condition with a non-trivial dependence on the space-time rapidity $\eta_s$. We apply our improved setup to provide predictions for the directed flow $v_1$ of $D$ mesons as a function of rapidity and for the nuclear modification factor $R_{\rm AA}$ and elliptic flow $v_2$ of muons from semileptonic decays of charmed and beauty hadrons at forward (and central) rapidity. In particular the HF directed flow turns out to be an interesting probe for three main reasons. First of all, due to the mismatch between the initial location of the HQ's production points and the bulk matter distribution of the fireball, it allows one to perform a full three-dimensional tomography of the medium~\cite{Chatterjee:2017ahy}: this by the way leads to a $v_1$ signal for $D$ mesons much larger then the one of light hadrons, which are thermally produced at freeze-out when the expanding fluid gets converted into a system of free-streaming particles decoupling from the latter.
Secondly, extending the theory-to-data comparison to the directed flow $v_1(y)$ of charm allows one to further constrain the HF transport coefficients. 
Finally, HQ's -- produced immediately after the heavy-ion collision -- witness the initial huge magnetic field present in the medium ($B\sim 10^{15}$ T). Any possible difference in the (electrically neutral) $D^0$ and $\overline{D^0}$ meson distributions (in particular the directed flow $v_1(y)$, the magnetic field being orthogonal to the reaction plane) can only arise from the partonic stage, in which charm quarks and antiquarks carry opposite electric charge and hence feel an opposite force once embedded in an electromagnetic field. This possibility was suggested in a few seminal papers~\cite{Das:2016cwd,Chatterjee:2018lsx} and people looked for evidences of such an effect in the experimental data. So far a coherent picture of STAR~\cite{Adam:2019wnk} and ALICE~\cite{Acharya:2019ijj} data at RHIC and at the LHC is missing. Before including the effect of the electromagnetic field, we believe it is important to correctly reproduce the size of the average ($v_1^{D+\overline D}$) signal, which is what we plan to do in our paper. In the next future we plan to go one step further, including the effect of the Lorentz force in the Langevin evolution of the HQ's.

Our paper is organized as follows. In Sec.~\ref{sec:hydro} we discuss our modelling of the background medium through which the propagation of the HQ's takes place, focusing in particular on the initial conditions of the hydrodynamic evolution.
In Sec.~\ref{sec:results} we present the results of our transport setup for various HF observables: in Sec.~\ref{sec:directed} we condider the directed flow of $D$ mesons, comparing our findings with recent results by the STAR and ALICE collaborations; in Sec.~\ref{sec:forward} we address the nuclear modification factor and the elliptic flow of HF particles in various rapidity intervals, comparing results in the forward and central rapidity regions.
Finally, in Sec.~\ref{sec:discussion} we discuss our major findings, providing an outlook of future improvements. 
     
\section{The 3+1 hydrodynamic background}\label{sec:hydro}
\begin{table}[!h]
\begin{center}
\begin{tabular}{|c|c|c|}
\hline
{} & Au-Au $\sqrt{s_{\rm NN}}\!=\!200$ GeV & Pb-Pb $\sqrt{s_{\rm NN}}\!=\!5.02$ TeV\\
\hline
$s_0$ (fm$^{-3}$) & 84 & 400\\
\hline
$\tau_0$ (fm/c) & 1 & 0.5\\
\hline
$\alpha_h$ & 0.15 & 0.15\\
\hline
$\eta_{\rm flat}$ & 1.5 & 1.5\\
\hline
$\sigma_\eta$ & 1 & 2.2 \\
\hline
$\eta_m$ & 3.36 ($=y_{\rm beam}\!-\!2$) & {8.58} ($=y_{\rm beam}$)\\
\hline
\end{tabular}
\end{center}
\caption{The parameters of the initial conditions employed to describe Au-Au and Pb-Pb collisions at RHIC and LHC center-of-mass energies.}
\label{tab:init}
\end{table}
\begin{figure}[!ht]
\begin{center}
\includegraphics[clip,width=0.48\textwidth]{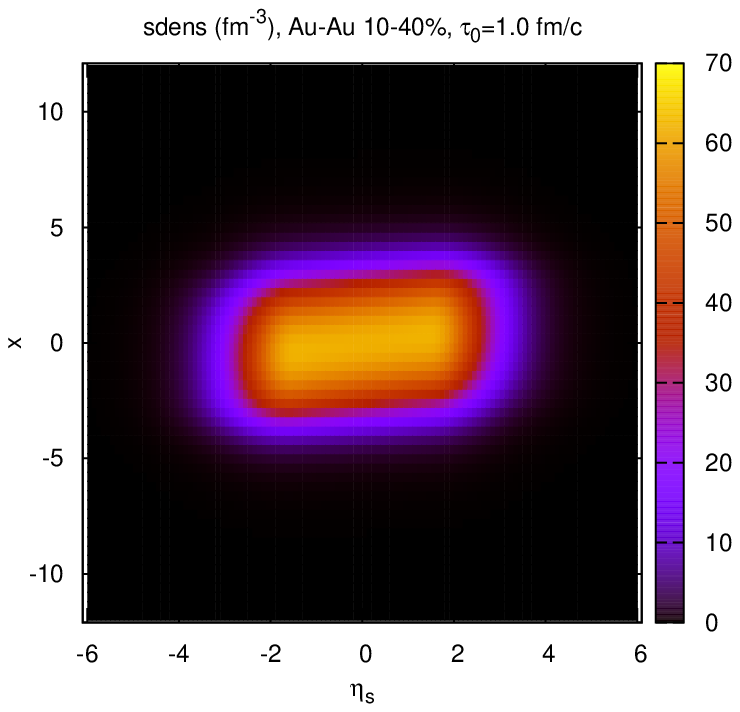}
\includegraphics[clip,width=0.48\textwidth]{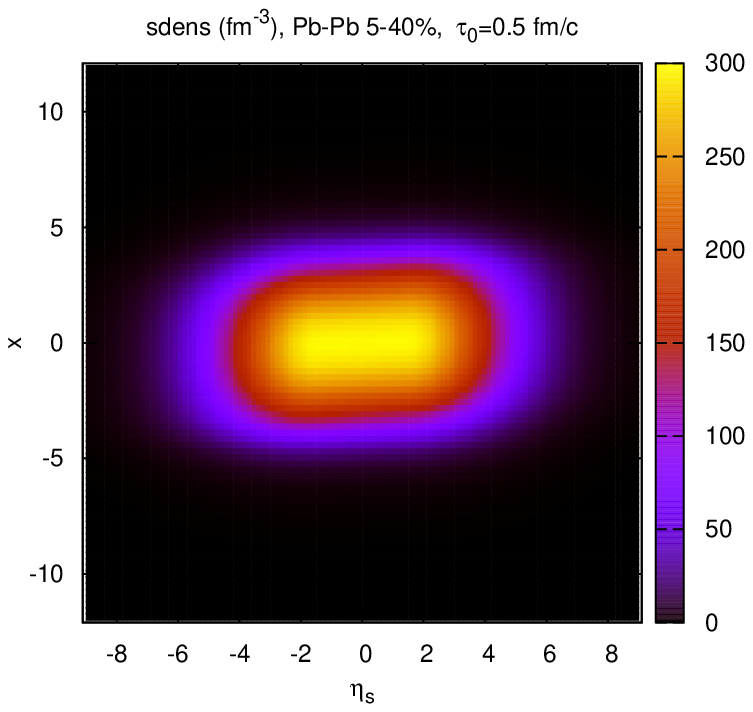}
\caption{The initial entropy density in the $\eta_s-x$ plane at $y=0$ in Au-Au collisions at $\sqrt{s_{\rm NN}}\!=\!200$ GeV (left panel) and Pb-Pb collisions at $\sqrt{s_{\rm NN}}\!=\!5.02$ TeV (right panel).}\label{fig:tilted}
\end{center}
\end{figure}
In order to set the initial conditions for our (3+1)D hydrodynamic equations following Ref.~\cite{Bozek:2010bi} we employ the optical Glauber model, assuming that the initial entropy deposition arises from a linear combination of the contributions from the participant (``wounded'') nucleons and from the binary nucleon-nucleon collisions, with relative weight $(1-\alpha_h)$ and $\alpha_h$, respectively. We assume that at the initial longitudinal proper time $\tau_0$ the fireball has a finite extension in rapidity ($\eta_s\equiv\frac{1}{2}\ln\frac{t+z}{t-z}$) modelled by the function
\beq
  H(\eta_s)=\exp\left[-\dfrac{(|\eta_s|-{\eta_{\rm flat}})^2}{2{\sigma_{\eta}}^2}\theta(|\eta_s|-{\eta_{\rm flat}})\right],
\eeq
with a central flat plateau of extension 2$\eta_{\rm flat}$ beyond which the density drops to zero with a Gaussian smearing $\sigma_\eta$. Furthermore right (left)-moving wounded nucleons are assumed to produce more particles at forward (backward) rapidity, respectively. The effect is parametrized by the function
\beq
f_{+/-}(\eta_s)=
\begin{cases}
0/2 & \eta_s < -\eta_m\\
\dfrac{\pm\eta_s+{\eta_m}}{{\eta_m}} & -\eta_m \le \eta_s \le \eta_m\\
2/0 & \eta_s > \eta_m.
\end{cases}
\eeq
The initial entropy density at $\tau_0$ is then given by
\beq
s(\x,\eta_s;b)=s_0\frac{(1\!-\!\alpha_h)[n_{\rm part}^A(\x;b)f_+(\eta_s)\!+\!n_{\rm part}^B(\x;b)f_-(\eta_s)]+\alpha_h n_{\rm coll}(\x;b)}{(1\!-\!\alpha_h)n_{\rm part}(\0;0)+\alpha_h n_{\rm coll}(\0;0)}H(\eta_s).\label{eq:entropy}
\eeq 
In the above equation the different modulation in rapidity of the contribution of participant nucleons from nuclei $A$ and $B$ is responsible for the tilting of the fireball, which eventually gives rise to the non-vanishing directed flow of the emitted hadrons. Furthermore, the tilted fireball is characterized by a sizable angular momentum and vorticity of the velocity field~\cite{Becattini:2015ska}, which could explain the non-zero polarization of $\Lambda$ baryons and vector mesons recently observed at RHIC~\cite{STAR:2017ckg} and at the LHC~\cite{Acharya:2019vpe}.

Concerning the initial condition for the fluid velocity, consistently with the pioneering study of Ref.~\cite{Bozek:2010bi}, we assume at longitudinal proper time $\tau_0$ a Bjorken-like flow with $V_z\!=\!z/t$ and $V_x\!=\!V_y\!=\!0$, neglecting any initial transverse expansion. Hence, at the beginning of the hydrodynamic evolution, the fluid rapidity $Y\!\equiv\!{\rm atanh}(V_z)$ coincides with the spatial rapidity  $\eta_s\!\equiv\!{\rm atanh}(z/t)$. Later on the non-vanishing pressure gradient along the longitudinal axis provides an acceleration which makes $Y\!>\!\eta_s$ for $\tau\!>\!\tau_0$. More complex initial flow profiles can be explored in the future.

In this paper we consider Au-Au and Pb-Pb collisions at $\sqrt{s_{\rm NN}}=200$ GeV and 5.02 TeV, respectively. In Table~\ref{tab:init} we summarize the parameters employed to initialize the hydrodynamic evolution. In Fig.~\ref{fig:tilted} we display the initial entropy density in the $\eta_s-x$ plane for non-central Au-Au and Pb-Pb collisions at RHIC and LHC center-of-mass energies. We notice the milder tilting of the fireball at LHC energy necessary in order to reproduce the experimental data.
Employing the values given in Table~\ref{tab:init} for the overall factor $s_0$ in Eq.~(\ref{eq:entropy}), corresponding to the initial entropy density at the center of the fireball for a collision at zero impact parameter ($b\!=\!0$), after integrating over the transverse plane at $\tau=\tau_0$ we obtain an entropy per unit rapidity $dS/d\eta_s\approx 14600$ for central ($0-5\%$) Pb-Pb collisions at the LHC. This quantity, neglecting dissipative effects during the expansion, is directly related to the final rapidity density of charged hadrons measured by the detectors.
Notice that, in case we assumed an exact scaling of $s(\x)$ with $n_{\rm coll}(\x)$ setting $\alpha_h=1$ in Eq.~(\ref{eq:entropy}) keeping $s_0$ untouched, we would get a lower value for the initial total entropy $dS/d\eta_s\approx 12000$. This is due to the fact that the distribution of binary nucleon-nucleon collisions in the transverse plane is narrower than the one of participant nucleons and hence the integration covers a smaller spatial domain.
\begin{figure}[!ht]
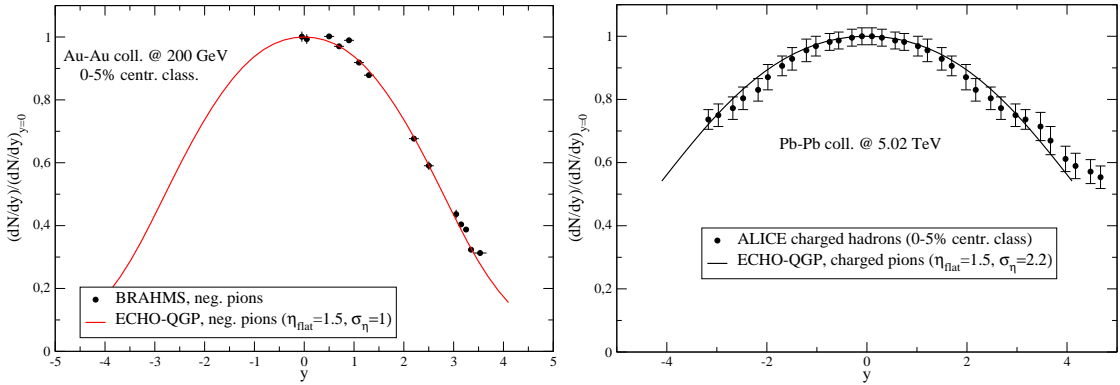

\begin{center}
\includegraphics[clip,width=0.48\textwidth]{pion-dNdy_BRAHMS.eps}
\includegraphics[clip,width=0.48\textwidth]{pion-dNdy-ALICE.eps}
\caption{The rapidity density of charged pions in the 0-5\% most central Au-Au collisions at $\sqrt{s_{\rm NN}}\!=\!200$ GeV (left panel) and Pb-Pb collisions at $\sqrt{s_{\rm NN}}\!=\!5.02$ TeV (right panel). Results obtained with the ECHO-QGP code starting from the initial conditions in Table~\ref{tab:init} are compared to experimental results obtained by the BRAHMS~\cite{Bearden:2004yx} and ALICE~\cite{Adam:2016ddh} collaborations at RHIC and at the LHC, respectively. ALICE data actually refer to the pseudo-rapidity density.}\label{fig:dNdy-soft}
\end{center}
\end{figure}
\begin{figure}[!ht]
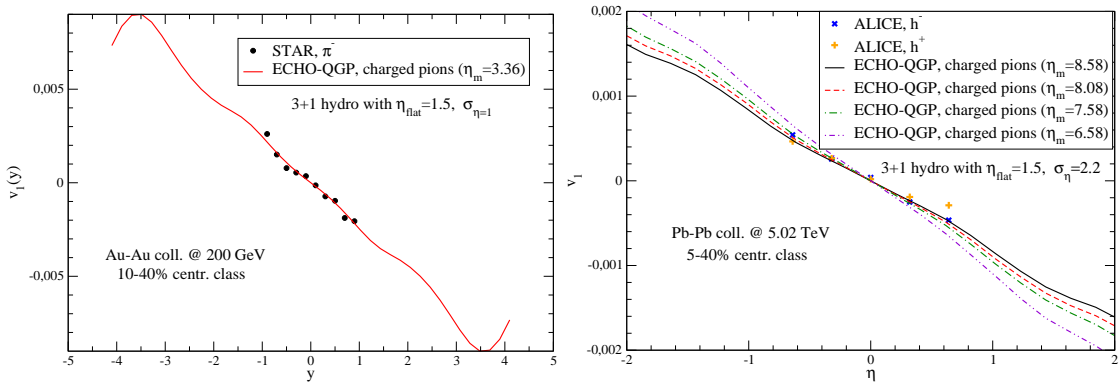

\begin{center}
\includegraphics[clip,width=0.48\textwidth]{pion-v1-STAR_10-40.eps}
\includegraphics[clip,width=0.48\textwidth]{pion-v1-ALICE.eps}
\caption{The directed flow $v_1(y)$ of charged pions in Au-Au collisions at $\sqrt{s_{\rm NN}}\!=\!200$ GeV (left panel) and Pb-Pb collisions at $\sqrt{s_{\rm NN}}\!=\!5.02$ TeV (right panel). Results obtained with the ECHO-QGP code starting from the initial conditions in Table~\ref{tab:init} are compared to experimental results obtained by the STAR~\cite{Adamczyk:2014ipa} and ALICE~\cite{Acharya:2019ijj} collaborations at RHIC and at the LHC, respectively.}\label{fig:v1-soft}
\end{center}
\end{figure}
\begin{figure}[!ht]
\begin{center}
\includegraphics[clip,width=0.48\textwidth]{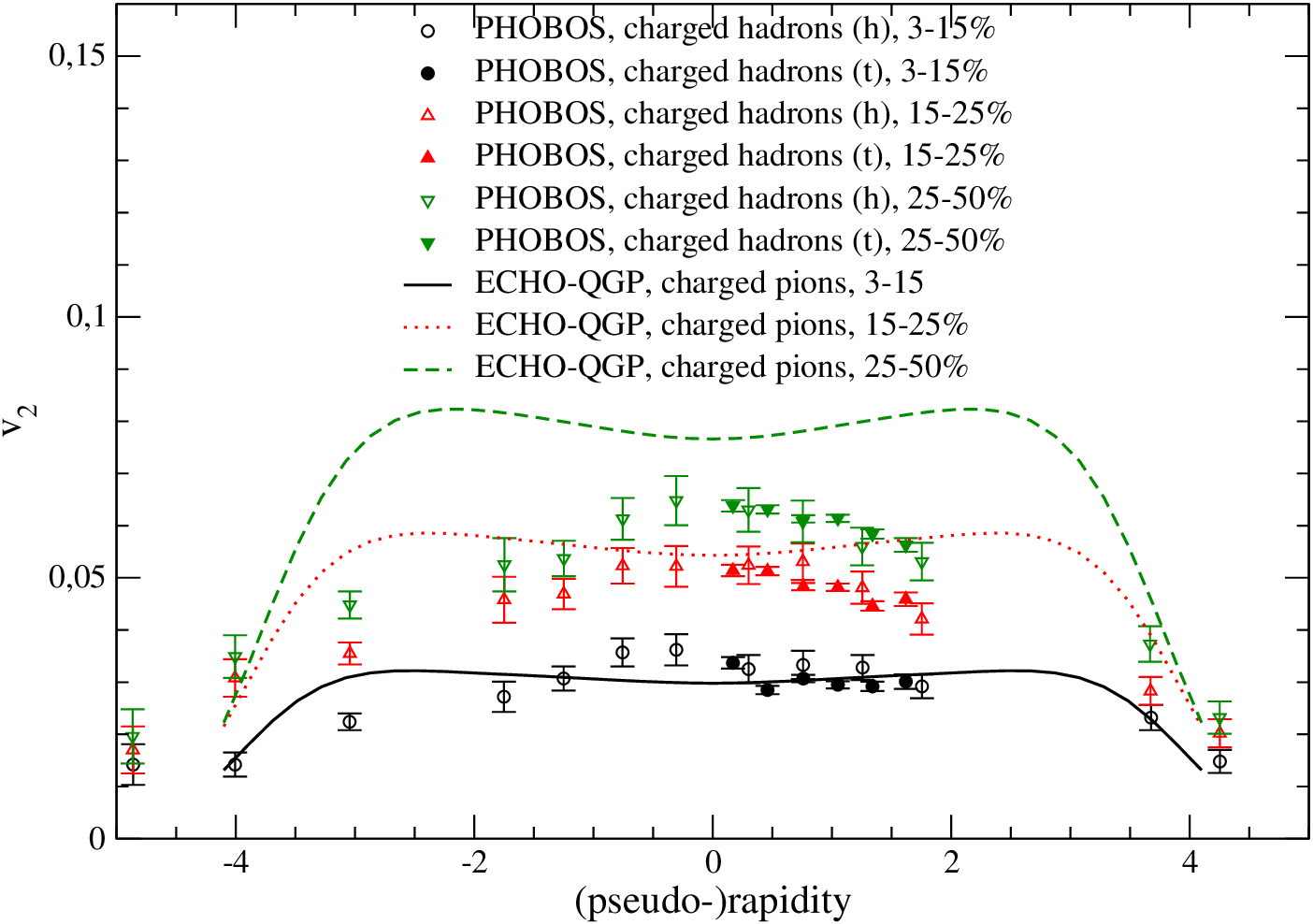}
\includegraphics[clip,width=0.48\textwidth]{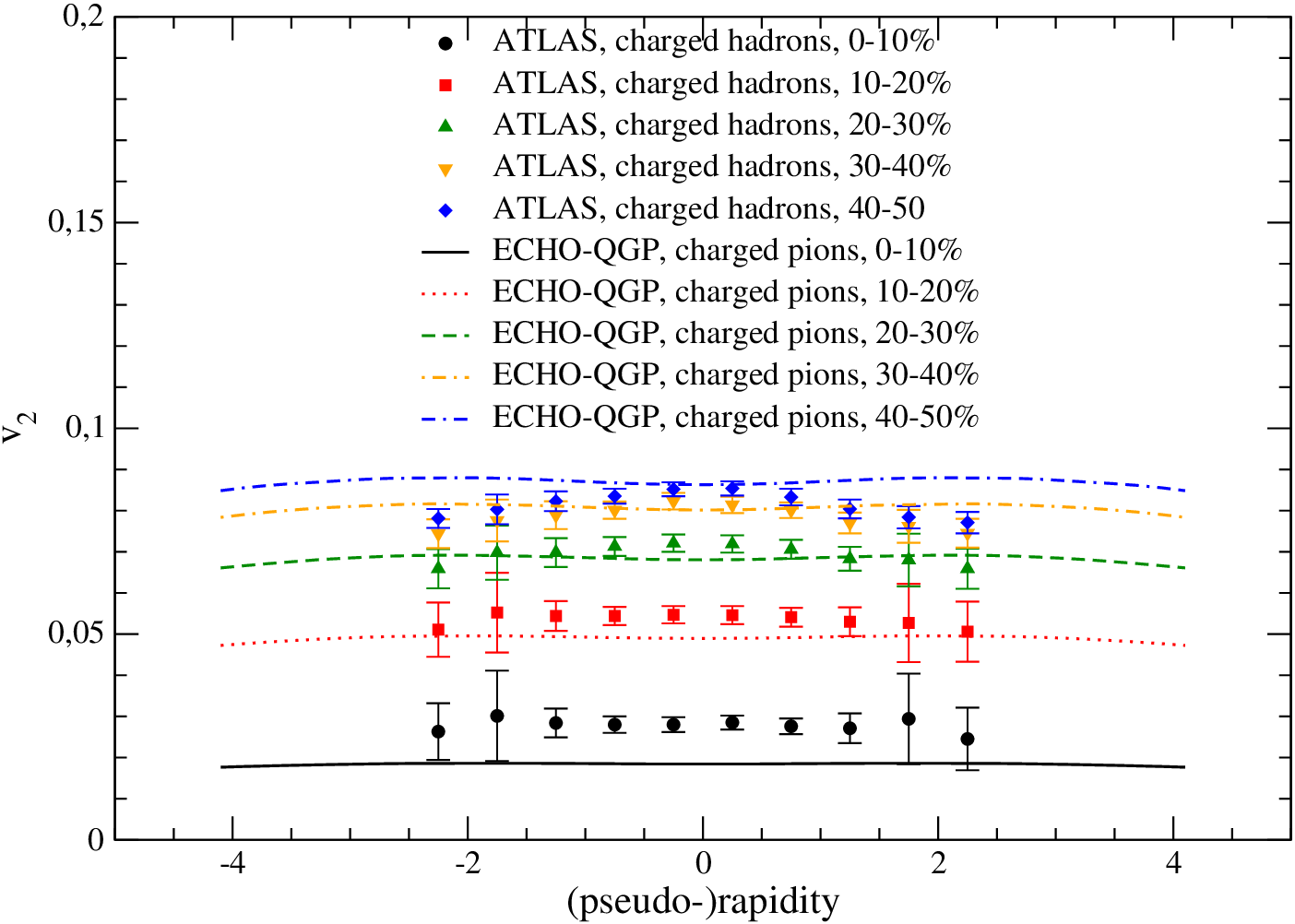}
\caption{The elliptic flow $v_2(y)/v_2(\eta)$ of charged pions/hadrons in Au-Au collisions at $\sqrt{s_{\rm NN}}\!=\!200$ GeV (left panel) and Pb-Pb collisions at $\sqrt{s_{\rm NN}}\!=\!5.02$ TeV (right panel). Results obtained with the ECHO-QGP code starting from the initial conditions in Table~\ref{tab:init} are compared to experimental results obtained by the PHOBOS~\cite{Back:2004mh} and ATLAS~\cite{Aad:2014eoa} collaborations at RHIC and at the LHC, respectively. PHOBOS data are obtained with two different methods depending on the rapidity interval: hits (h) and tracks (t). ATLAS data actually refer to $\sqrt{s_{\rm NN}}\!=\!2.76$ TeV.}\label{fig:v2-soft}
\end{center}
\end{figure}

Before performing our transport simulations to describe HF observables, here we validate our hydrodynamic background against experimental data for soft particle production. We do not aim at performing a global precision fit, but simply to show that we are able to provide a reasonable description of the collective expansion of the fireball employing the parameter set in Table~\ref{tab:init}. For the kinetic decoupling temperature for soft hadrons we set the value $T_{\rm FO}\!=\!140$ MeV, common to all centrality classes.
In Fig.~\ref{fig:dNdy-soft} we show the rapidity density of soft hadrons. We compare ECHO-QGP results for primary pions (i.e. excluding the feed-down contribution from resonance decay) to the experimental data obtained in central (0-5\%) Au-Au and Pb-Pb collisions at RHIC and at the LHC: negative pions measured by the BRAHMS experiment (left panel,~\cite{Bearden:2004yx}) and charged hadrons measured by the ALICE collaboration (right panel,~\cite{Adam:2016ddh}).
In Fig.~\ref{fig:v1-soft} we focus on the directed flow of soft particles at RHIC and at the LHC. We compare the results for primary pions obtained with our hydrodynamic background to the directed flow of negative pions and charged hadrons measured in non-central Au-Au and Pb-Pb collisions by the STAR and ALICE collaborations. We notice that the directed flow of soft particles is larger at lower center-of-mass energy. In order to reproduce the latter, for the parameter describing the initial tilting of the fireball we have to fix the values $\eta_m=y_{\rm beam}-2=3.36$ at $\sqrt{s_{\rm NN}}=200$ GeV and $\eta_m=y_{\rm beam}=8.58$ at $\sqrt{s_{\rm NN}}=5.02$ TeV.
Finally, in Fig.~\ref{fig:v2-soft} we display the rapidity dependence of the elliptic flow of soft hadrons for various centrality classes. Hydrodynamic predictions for primary pions obtained with the ECHO-QGP code starting from the initial conditions summarized in Table~\ref{tab:init} are compared to experimental data for the $v_2(\eta)$ ($\eta$ being the pseudorapidity) of charged particles collected by the PHOBOS and ATLAS collaborations in Au-Au and Pb-Pb collisions at RHIC and LHC, respectively. Notice that ATLAS data refer to  $\sqrt{s_{\rm NN}}=2.76$ TeV. Overall the agreement is quite satisfactory, even if we tend to overestimate the $v_2$ in the most peripheral collisions and to get a flatter behaviour for a larger rapidity range as compared to the experimental data, which (in particular at RHIC) tend to decrease linearly for increasing rapidity. We also slightly underestimate the elliptic flow in the most central collisions at the LHC, but this can be attributed to our optical-Glauber initial conditions, which miss event-by-event eccentricity fluctuations.

In summary, our modelling of the bulk medium produced in heavy-ion collisions allows us to get a quite satisfactory description of several soft observables and can be used as a reasonable background to study the propagation of HQ's. Our description of the fireball can be surely improved in the next future, exploring more realistic profiles for the initial longitudinal velocity of the fluid than the current Bjorken ansatz $V_z=z/t$, adopting a MC-Glauber initialization of the density of the medium also in the study of the directed flow and including all soft hadrons and resonances in the validation against the experimental data. 
\section{Heavy-flavour transport simulations}\label{sec:results}
We now address the issue of the propagation of HQ's in the tilted background described in the previous section. As discussed in detail in previous publications the transport of charm and beauty quarks in the QGP is described through {the following relativistic Langevin equation~\cite{Beraudo:2014boa,Beraudo:2017gxw}
\beq
{\Delta \vec{p}}/{\Delta t}=-{\eta_D(p)\vec{p}}+{\vec\xi(t)}\label{eq:Langevin}
\eeq
containing a deterministic friction force quantified by the drag coefficient $\eta_D$ and a random noise term specified by its temporal correlator
\beq
\langle\xi^i(\vec p_t)\xi^j(\vec p_{t'})\rangle\!=\!{b^{ij}(\vec p_t)}{\delta_{tt'}}/{\Delta t}\;,\qquad{b^{ij}(\vec p)}\!\equiv\!{\kk_\|(p)}\pp^i\pp^j+{\kk_\perp(p)}(\delta^{ij}\!-\!\pp^i\pp^j).
\eeq
Different values for the longitudinal/transverse momentum-diffusion coefficients $\kk_\|$ and $\kk_\perp$ are employed}, arising either from a weak-coupling calculation with resummation of medium effects (HTL scheme) or from lattice-QCD simulations (l-QCD scheme). Both schemes display some shortcomings due to the mismatch between the kinematic and temperature conditions of experimental relevance and the ones in which theoretical calculations can provide results based on solid grounds. {The drag coefficient $\eta_D$ is connected to the momentum-diffusion coefficients $\kk_\|$ and $\kk_\perp$ by a generalized Einstein relation, ensuring the asymptotic approach to kinetic equilibrium}.
{Hadronization is modelled recombining with probability 1 the heavy quarks with light thermal partons from the same fluid-cell (an instantaneous decoupling at temperature $T_H\!=\!155$ MeV, with no further rescattering in the hadron-gas phase is assumed), forming excited $Q\overline q$ (or $\overline Q q$) strings which are then fragmented in a $2\to 1^*\to N$ process according to the Lund model implemented in PYTHIA 6.4. This turns out to have a major effect on the final charm and beauty hadron distributions.} 
For further details we refer the reader to our past publications~\cite{Beraudo:2014boa,Beraudo:2017gxw}. 

It is important to stress that, at variance with the background fireball, the spatial distribution of HQ's at the initial thermalization time of the bulk medium $\tau_0$ does not display any distorsion in the $\eta_s-x$ plane. HQ's are in fact produced in hard scattering processes in binary nucleon-nucleon collisions, hence their initial location in the transverse plane simply follows the $n_{\rm coll}(\x)$ distribution (we do not consider any transverse expansion before $\tau_0$ neither for the medium nor, for consistency, for the HQ's); concerning their longitudinal position, neglecting any interaction with the medium before $\tau_0$, one has simply $z=v_zt$ ($v_z$ being the HQ velocity) and hence the spatial rapidity of their fluid cell coincides with the rapidity with which they are produced, namely
\beq
\eta_s=\frac{1}{2}\log\frac{t+z}{z-z}=\frac{1}{2}\log\frac{1+v_z}{1-v_z}=\frac{1}{2}\log\frac{E+p_z}{E-p_z}=y,
\eeq 
with no forward/backward asymmetry. Thus, in the case of HF particles, we have a second possible source of directed flow $v_1$ beside the anisotropic pressure gradients, i.e. the mismatch between the initial spatial distribution of the parent HQ's and the one of the bulk matter produced in the collision. In Sec.~\ref{sec:directed} we will investigate the quantitative consequences of such an occurrence, while in Sec.~\ref{sec:forward} we will profit from our (3+1)D setup to provide predictions for HF observables at forward rapidity, without the necessity of relying on the approximation of a longitudinal boost-invariant expansion.

\subsection{Directed flow and HF transport coefficients}\label{sec:directed}
\begin{figure}[!ht]
\begin{center}
\includegraphics[clip,width=0.48\textwidth]{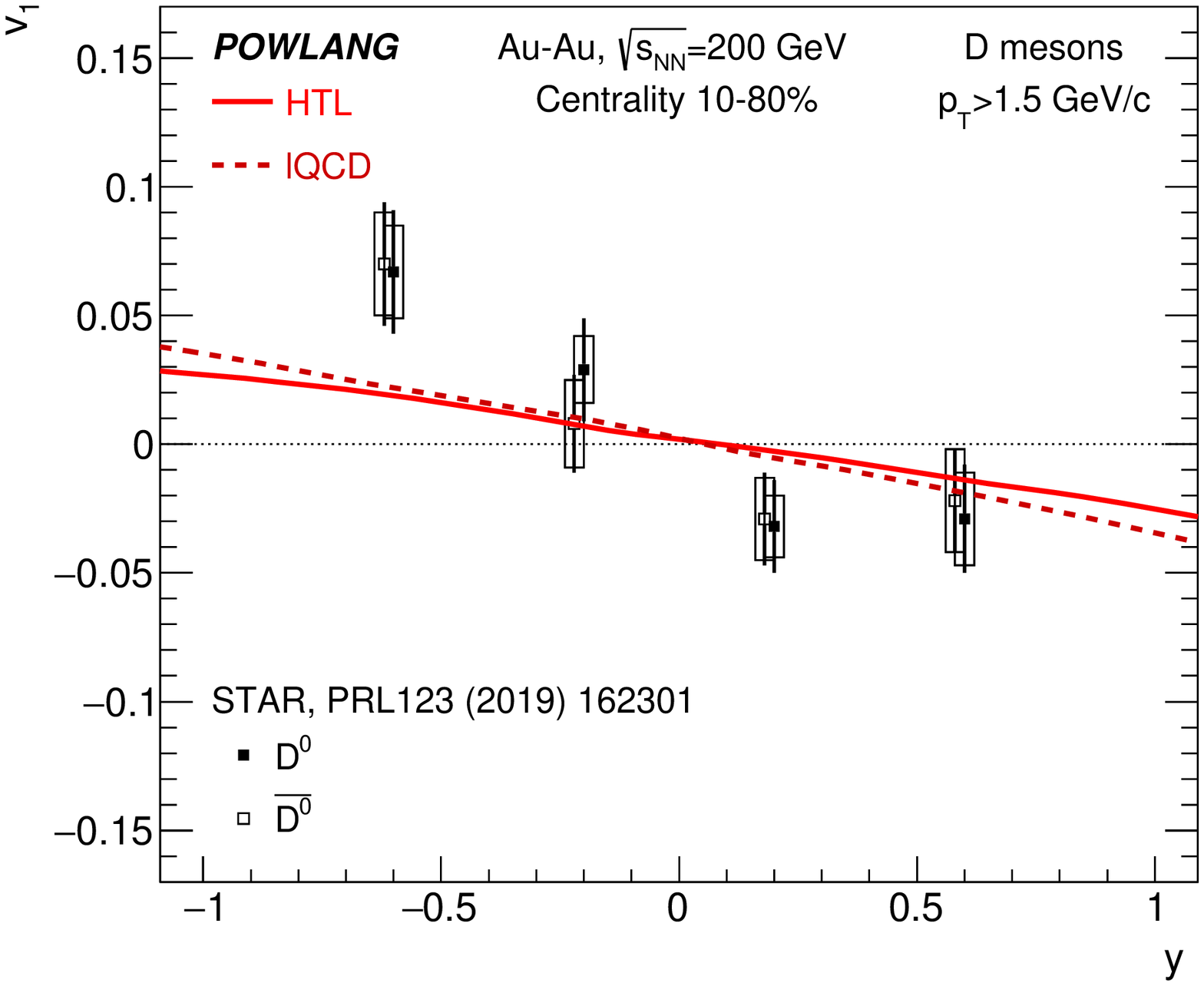}
\includegraphics[clip,width=0.48\textwidth]{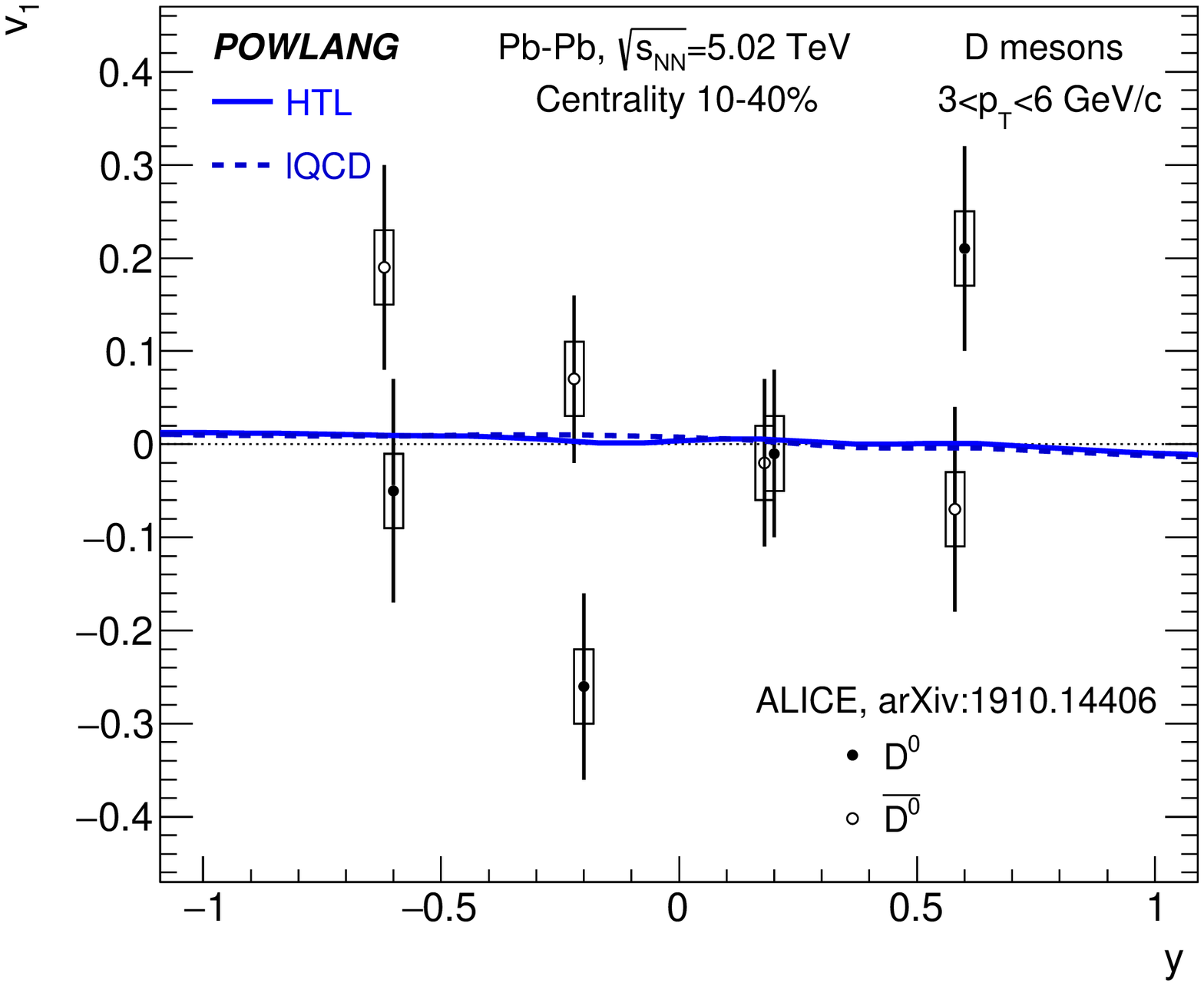}
\caption{The directed flow $v_1(y)$ of $D$ mesons in Au-Au collisions at $\sqrt{s_{\rm NN}}\!=\!200$ GeV (left panel) and Pb-Pb collisions at $\sqrt{s_{\rm NN}}\!=\!5.02$ TeV (right panel). POWLANG results for the average $D^0+\overline{D^0}$ flow with different different choices of transport coefficients are compared to experimental data for $D^0$ and $\overline{D^0}$ obtained by the STAR~\cite{Adam:2019wnk} and ALICE~\cite{Acharya:2019ijj} collaborations.}\label{fig:v1-D-vs-exp}
\end{center}
\end{figure}
In this section we focus on the study of the directed flow $v_1$ of $D$ mesons in non-central nucleus-nucleus collisions and on the information on the initial conditions and on the values of the transport coefficients that one can extract combining this measurement with the ones of radial and elliptic flow.
In Fig.~\ref{fig:v1-D-vs-exp} we display the predictions for the $D$-meson $v_1$ obtained with our transport setup, comparing our results to the data collected by the STAR and ALICE collaborations in Au-Au and Pb-Pb collisions at $\sqrt{s_{\rm NN}}\!=\!200$ GeV and $\sqrt{s_{\rm NN}}\!=\!5.02$ TeV, respectively. As one can see our results can qualitatively reproduce the trend of the data measured at RHIC (left panel), where the STAR experiment obtained evidence for a non-vanishing $v_1$ in Au-Au collision, with no significant difference between $D^0$ and $\overline{D^0}$. The situation is less clear for Pb-Pb collisions at the LHC, where we obtained a much weaker signal than at RHIC (consistent with the milder tilting of the fireball), but for which the ALICE collaboration found a significant $\Delta v_1$ between $D^0$ and $\overline{D^0}$ mesons, although the results are affected by large statistical and systematic uncertainties which prevent one from drawing firm conclusions. Since neither $D^0$ nor $\overline{D^0}$ mesons carry electric charge and hence their interaction with a hot hadron gas should be the same even in the presence of a non-vanishing electromagnetic field, the only conceivable origin of a different behaviour should be looked for in the response of the parent charm quarks/antiquarks to the strong primordial magnetic field generated by the spectator protons and partially frozen in the fireball if the QGP is characterized by a large electric conductivity. Addressing such an issue and understanding the origin of the diffent size of $\Delta v_1$ observed at RHIC and at the LHC would require coupling our transport setup to a full set of RMHD equations, providing a consistent description of the evolution of the matter and of the electromagnetic fields. We leave this for future work, waiting also for experimental data affected by smaller statistical and systematic uncertainties. In the following our results always refer to the average of particle and antiparticle contributions.

\begin{figure}[!ht]
\begin{center}
\includegraphics[clip,width=0.48\textwidth]{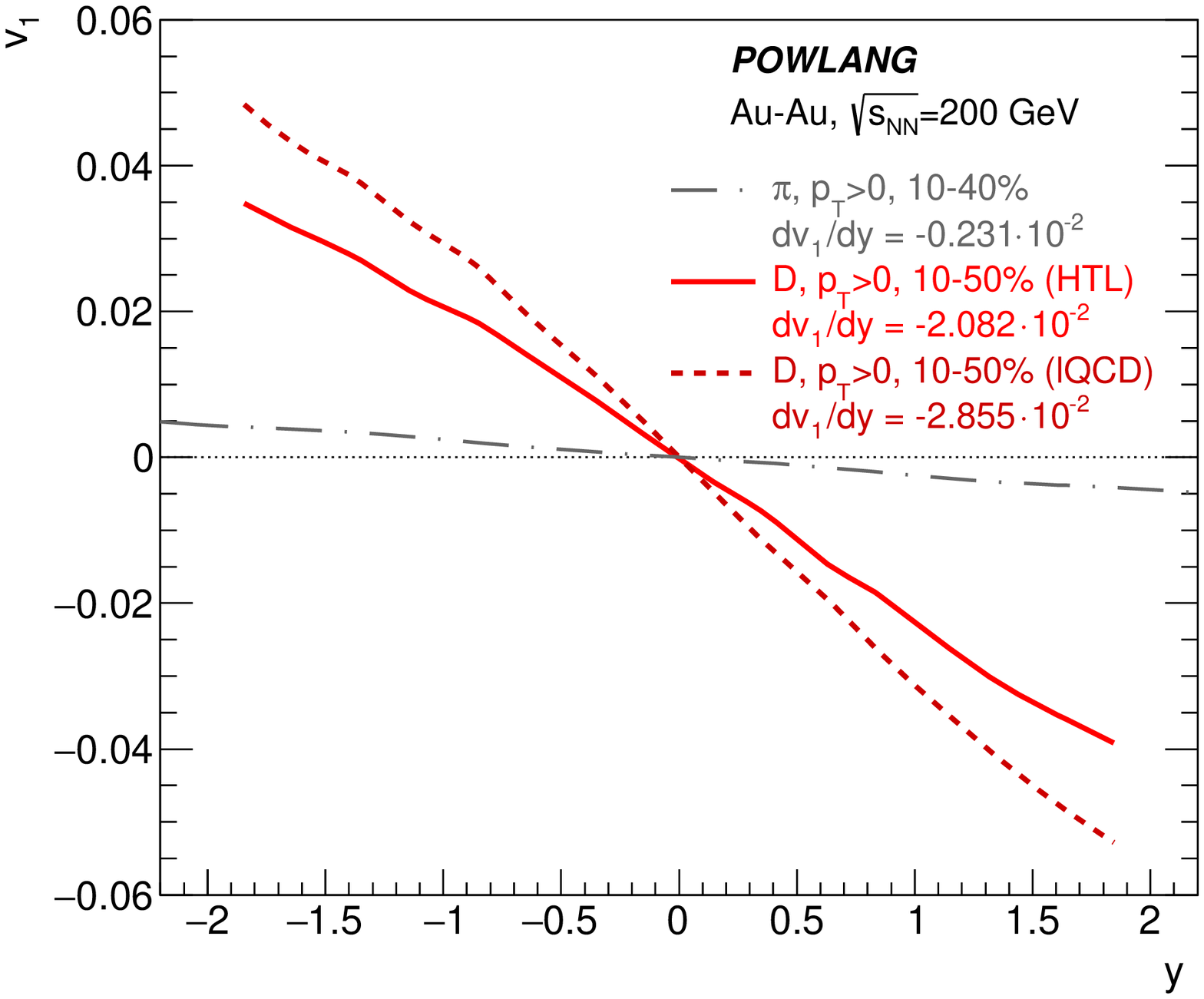}
\includegraphics[clip,width=0.48\textwidth]{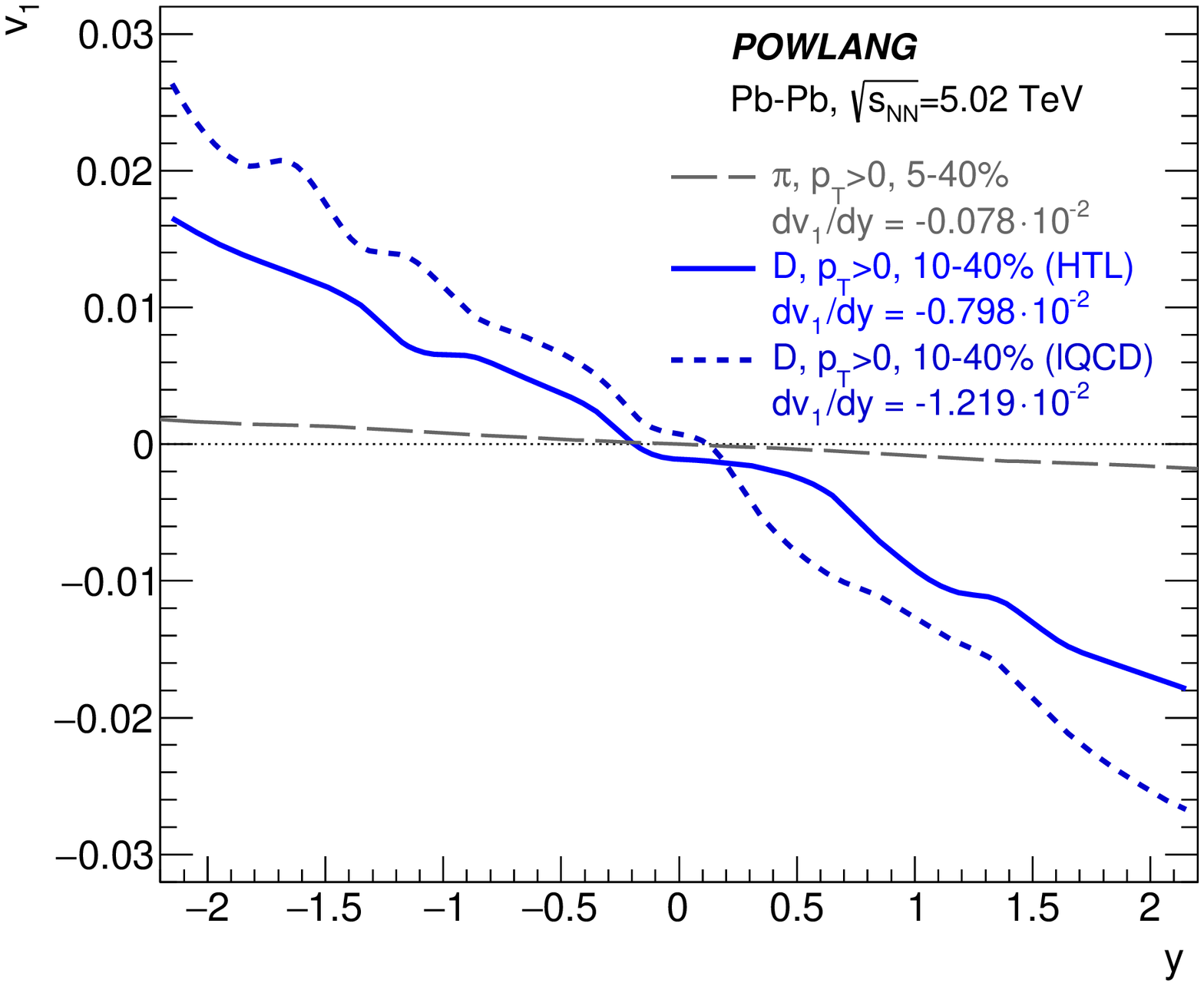}
\caption{The directed flow $v_1(y)$ of $D$ mesons  in Au-Au collisions at $\sqrt{s_{\rm NN}}\!=\!200$ GeV (left panel) and Pb-Pb collisions at $\sqrt{s_{\rm NN}}\!=\!5.02$ TeV (right panel) predicted by our (3+1)D transport setup compared to the one of primary charged pions provided by our hydrodynamic modelling of the background medium.}\label{fig:v1-D-vs-pi}
\end{center}
\end{figure}
As above discussed, the initial spatial distribution of the HQ's not matching the density of the bulk medium provides a second possible contribution to the $D$-meson $v_1$, besides the directed flow of the background arising from the initial tilting of the fireball: this may lead to a larger $v_1$ for $D$ mesons than for light hadrons. We perform such a comparison in Fig.~\ref{fig:v1-D-vs-pi}.
As one can see, the directed flow of charmed hadrons quantified by the slope $|dv_1/dy|$ turns out to be about one order of magnitude larger than the one of primary pions; both at RHIC and at LHC energies one gets a larger $v_1$ for $D$ mesons with lattice-QCD than HTL transport coefficients.  

\begin{figure}[!ht]
\begin{center}
\includegraphics[clip,width=0.48\textwidth]{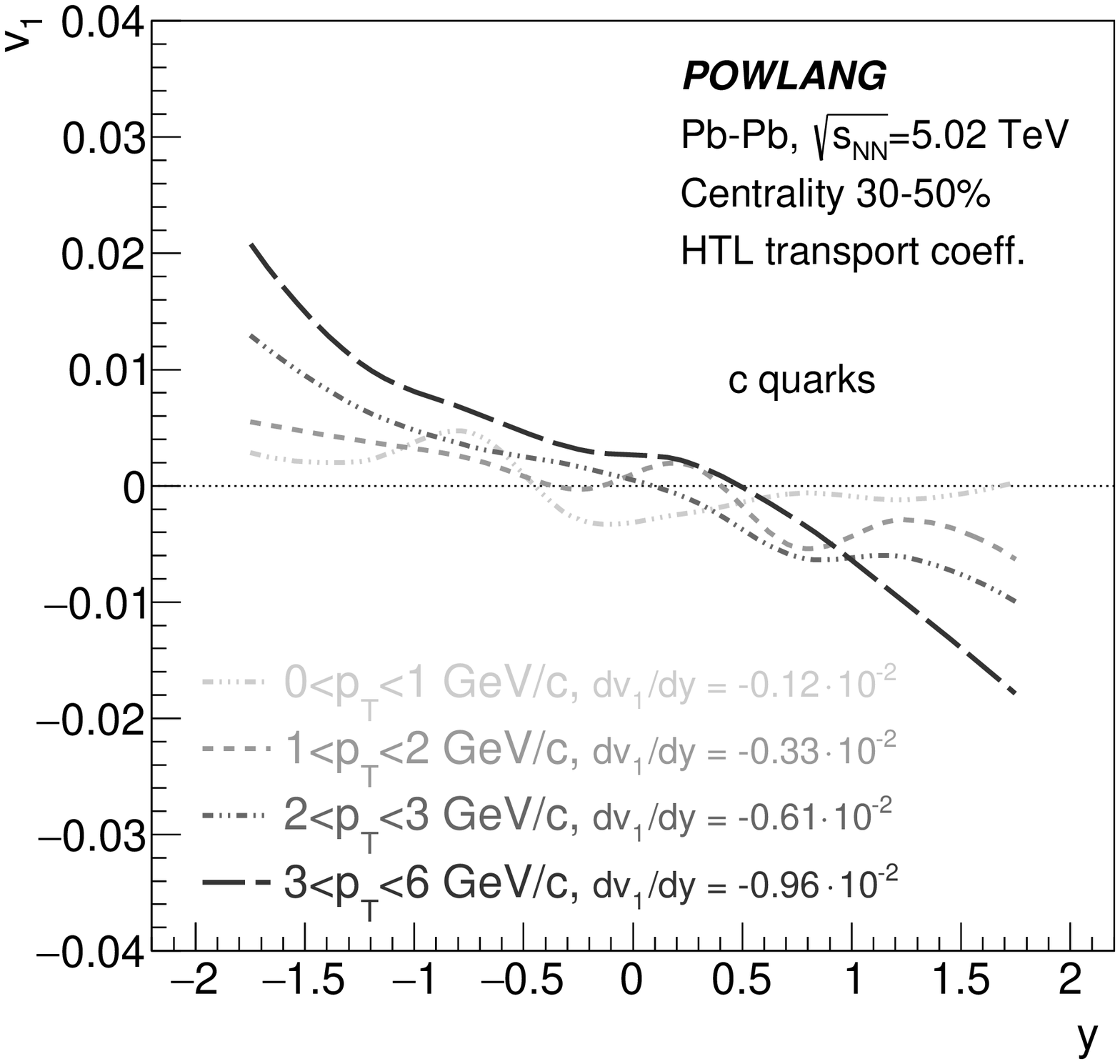}
\includegraphics[clip,width=0.48\textwidth]{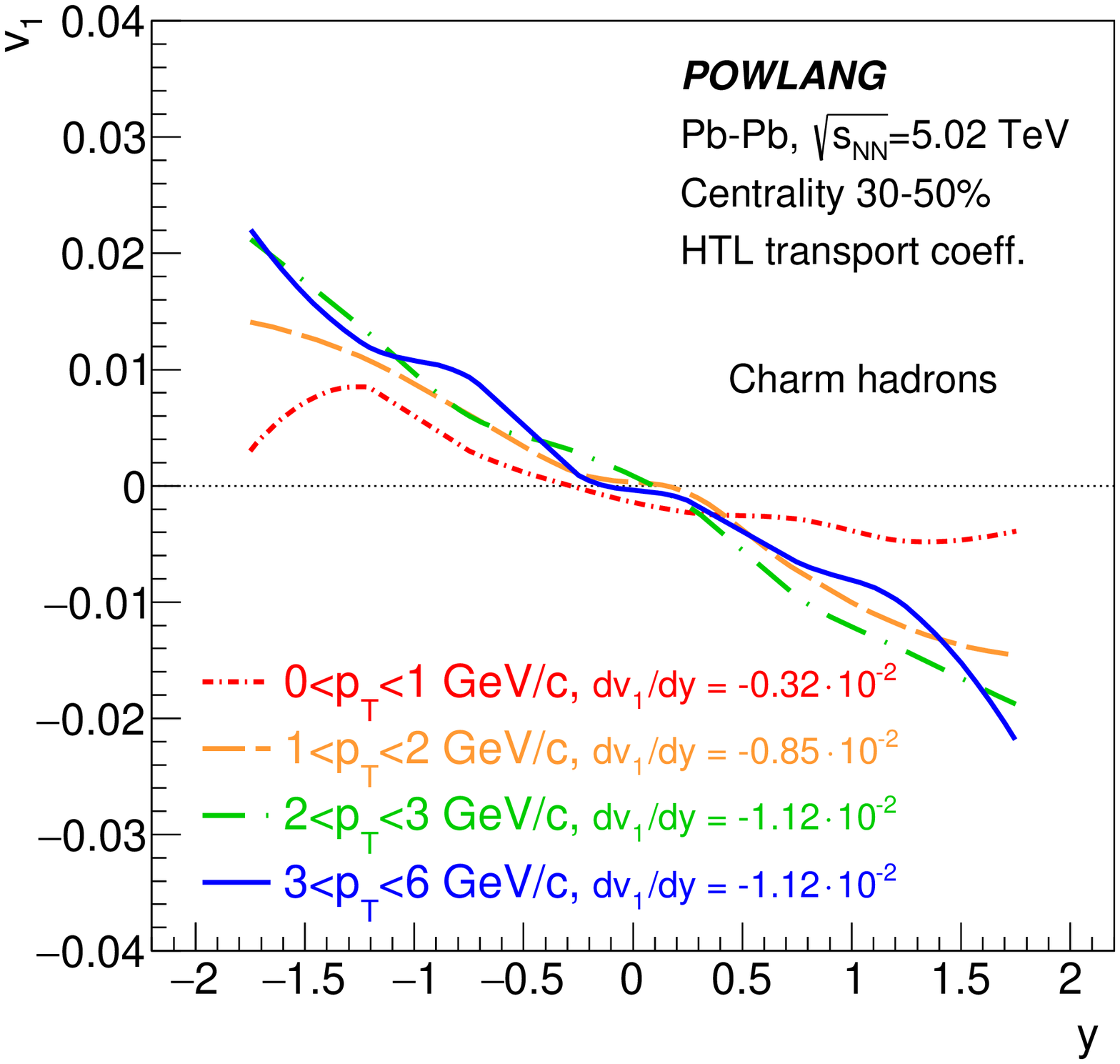}
\caption{The directed flow $v_1(y)$ of charm quarks (left panel) and $D$ mesons (right-panel) in non-central Pb-Pb collisions at $\sqrt{s_{\rm NN}}\!=\!5.02$ TeV for different $p_T$ bins. {Predictions with HTL transport coefficients are displayed}. For $p_T\lsim 3$ GeV/c hadronization via recombination with light thermal partons from the same fluid cell enhances the final $v_1(y)$.}\label{fig:v1-c-vs-D}
\end{center}
\end{figure}
Hadronization was found to significantly affect the final momentum and angular distributions of charmed and beauty hadrons. Recombination of HQ's with light thermal partons makes HF hadrons inherit part of the collective flow of the medium, which turns out to be important in order to get values of the $v_2$ and $v_3$ coefficients in agreement with the experimental data. In Fig.~\ref{fig:v1-c-vs-D} we show how also the directed flow is affected by hadronization. For $p_T\lsim 3$ GeV/c the recombination of charm quarks with light thermal partons from the same fluid cell (which in POWLANG gives rise to the formation of colour-singlet strings to further fragment) enhances the $v_1(y)$ of the final $D$ mesons. If one considers the very low value of the $v_1$ of charged hadrons this may appear surprising, since the overall momentum distribution of light quarks should accordingly display a very small dipole asymmetry. However, HQ's do not perform a uniform sampling of the hadronization hypersurface at $T=T_c$ and of the momenta of the thermal partons with which they recombine: there is a strong correlation between the momentum of the HQ's and the position and velocity of the fluid cell in which they are found at $T=T_c$ and hence the momentum of the light quarks with which recombination occurs. This explains why in any case the directed flow of $D$ mesons is larger than the one of the parent charm quarks.   

\begin{figure}[!ht]
\begin{center}
\includegraphics[clip,width=0.99\textwidth]{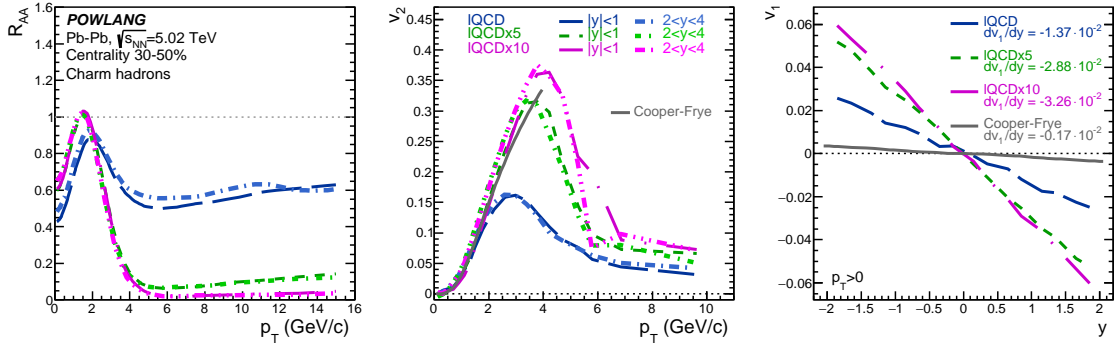}
\caption{The discriminating power of a combined measurement of the $R_{\rm AA}$ (left panel), $v_2$ (middle panel) and $v_1$ (right panel). POWLANG predictions for charm hadrons obtained with l-QCD tranport coefficients multiplied by 1, 5 and 10 are compared. We also display in grey the Cooper-Frye results for $D$-mesons in full thermodynamic equilibrium.}\label{fig:high-kappa-D}
\end{center}
\end{figure}
\begin{figure}[!ht]
\begin{center}
\includegraphics[clip,width=0.99\textwidth]{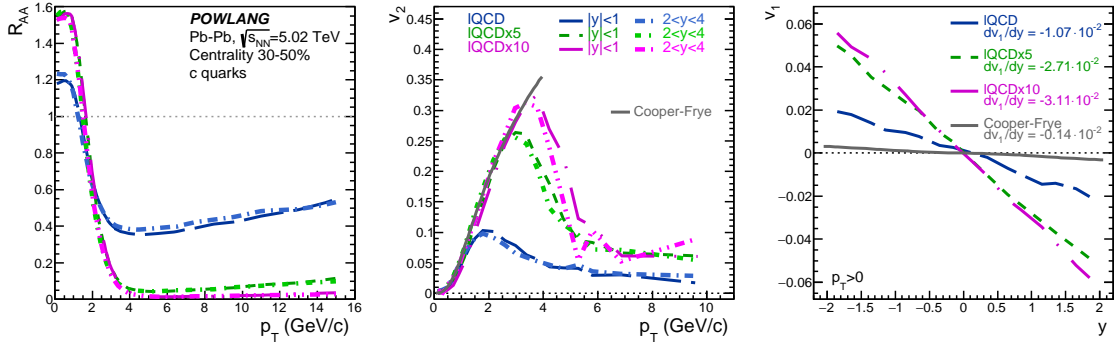}
\caption{The same as in Fig.~\ref{fig:high-kappa-D}, but for charm quarks. For the elliptic flow transport results for very large values of $\kappa$ are even closer to the hydrodynamic limit.}\label{fig:high-kappa-c}
\end{center}
\end{figure}
To what extent adding as an observable the directed flow $v_1(y)$ allows one to put tighter constraints on the value of the HF transport coefficients in the QGP? Clearly, with the same set of coefficients one must reproduce both the momentum distribution --  and specifically its nuclear modification factor $R_{\rm AA}$ -- and the various azimuthal flow harmonics $v_2$, $v_3$ (the latter not addressed here, since it would require dealing with event-by-event eccentricity fluctuations) and also $v_1$. It is interesting to notice that the twofold origin of the directed flow of charm hadrons leads to a non-trivial behaviour. In fact, one expects that as the HQ momentum-diffusion coefficient $\kappa$ increases HF particles should approach local thermal equilibrium with the fireball and hence their final momentum distribution should coincide with the one predicted by hydrodynamics and provided by a standard Cooper-Frye algorithm describing sudden particle decoupling from a freeze-out hypersurface.
Starting from the lattice-QCD result for $\kappa$ and multiplying the latter by a factor 5 and 10, in Figs.~\ref{fig:high-kappa-D} and~\ref{fig:high-kappa-c} we display the resulting $R_{\rm AA}$ (left panels), $v_2$ (middle panels) and $v_1$ (right panels) of charm hadrons and quarks, respectively. We notice that for such large values of $\kappa$ the results of transport calculations look very similar up to moderate values of $p_T$. However, the situation is very different for elliptic and directed flow. In the case of $v_2$ transport results approach the Cooper-Frye hydrodynamic limit, also shown for comparison in Figs.~\ref{fig:high-kappa-D} and~\ref{fig:high-kappa-c} (in the case of charm quarks the matching is perfect up to $p_T\approx 4$ GeV/c). On the other hand, in the case of $v_1$, transport calculations at strong coupling lead to a completely different behaviour, with a much larger directed flow than the one predicted by hydrodynamics.
This may appear an inconsistency of the model, but it admits a quite natural explanation. As already mentioned, the initial spatial distribution of the $Q\overline Q$ pairs is completely fixed by the position of the primordial nucleon-nucleon collisions in the transverse plane and by the momentum with which the HQ's are produced. In particular, at variance with the density of the background medium, the $Q\overline Q$ distribution does not display any spatial tilting. Hence, at a given rapidity, there is a mismatch between the center-of-mass of the fireball and the one of the $Q\overline Q$ distribution and this represents an important contribution to the final $v_1$. Furthermore, one has for the spatial diffusion coefficient $D_s=2T^2/\kappa$: the larger $\kappa$, the smaller $D_s$. Since $\langle \vec x^2\rangle\sim 6D_st$, for very large $\kappa$ the HQ's undergo a very limited spatial diffusion and the asymmetry between their distribution and the one of the bulk medium tends to survive for the entire hydrodynamic expansion of the fireball. 

\begin{figure}[!ht]
\begin{center}
\includegraphics[clip,width=0.99\textwidth]{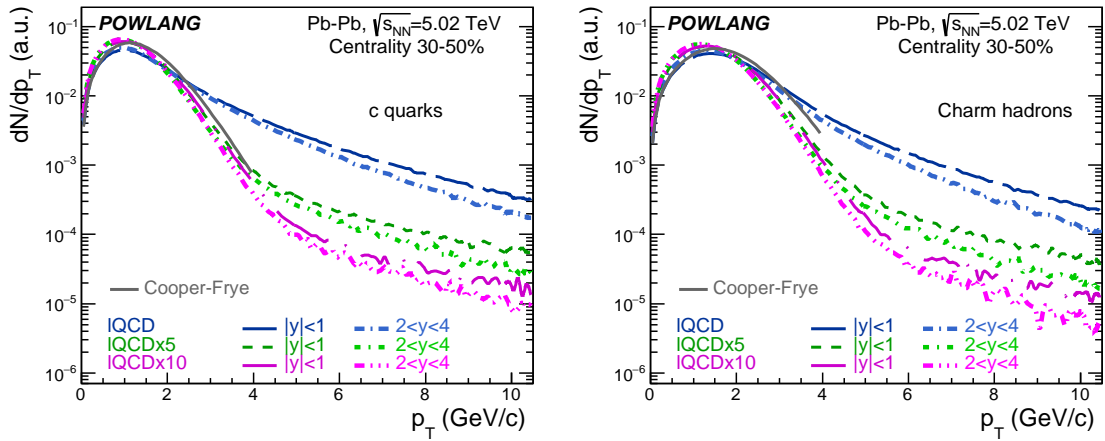}
\caption{A comparison of the transverse momentum spectra of charm quarks (left panel) and hadrons (right panel) predicted by transport calculations for different values of $\kappa$. We also display in grey the Cooper-Frye curves corresponding to the thermal emission of $c$ quarks and $D$ mesons from a decoupling hypersurface at $T=T_c=155$ MeV.}\label{fig:pt-spectra}
\end{center}
\end{figure}
Finally, in Fig.~\ref{fig:pt-spectra} we address the $p_T$-distribution of HF particles, both $c$ quarks (left panel) and charm hadrons (right panel). Also in this case transport outcomes for very large values of $\kappa$ tend to converge to the same result up to moderate $p_T$, but for the $D$ mesons the difference from the pure hydrodynamic prediction is slightly more pronounced than for charm quarks. The initial HQ distribution, following $n_{\rm coll}(\x)$, is narrower than the one of the background medium, whose initial density here follows a linear combination of  $n_{\rm coll}(\x)$ and $n_{\rm part}(\x)$. Due to the small value of $D_s$ the HQ spatial diffusion is limited, this HQ-bulk asymmetry survives and at decoupling HF particles perform a non-uniform sampling of the freeze-out hypersurface, with an overpopulation of HQ's decoupling from fluid cells closer to the origin and hence recombining with thermal partons with smaller collective radial velocity.

\subsection{Heavy-flavour at forward rapidity}\label{sec:forward}
Working with a full (3+1)D hydrodynamic background opens the possibility of addressing the study of HF observables at forward rapidity, relying on a more realistic description of the hot QCD medium than the one based on the assumption of longitudinal boost-invariance.

Measurements of open HF at forward rapidity were carried out so far only indirectly, via the detection of muons from the decay of charm and beauty hadrons through the muon spectrometer of the ALICE experiment. Results for the nuclear modification factor and the elliptic flow of HF decay muons were presented in Refs.~\cite{Abelev:2012qh,Adam:2015pga,Acharya:2020xzh} and compared to the predictions of some transport calculations~\cite{Uphoff:2012gb,He:2014cla,Nahrgang:2013xaa}. Unfortunately, so far it is not possible to separate the contributions from charm and beauty semileptonic decays, which will become accessible in Run 3 at the LHC thanks to the new Muon Forward Tracker~\cite{CERN-LHCC-2015-001}.
Important opportunities in Run 3 will arise also from the LHCb experiment, which after the upgrade of the detector will have the potential to measure $D$ and $B$ mesons at formard rapidity.
The second and third azimuthal harmonics of HF decay muons were also studied by the ATLAS experiment in a quite broad window $|\eta|<2$ around mid-rapidity~\cite{Aad:2020grf} and compared to various model predictions~\cite{Katz:2019fkc,Zigic:2018ovr}. The quite broad rapidity range covered by ATLAS measurements makes of interest having at disposal a theoretical setup non relying on the simplifying assumption of longitudinal boost-invariance.

\begin{figure}[!ht]
\begin{center}
\includegraphics[clip,width=0.48\textwidth]{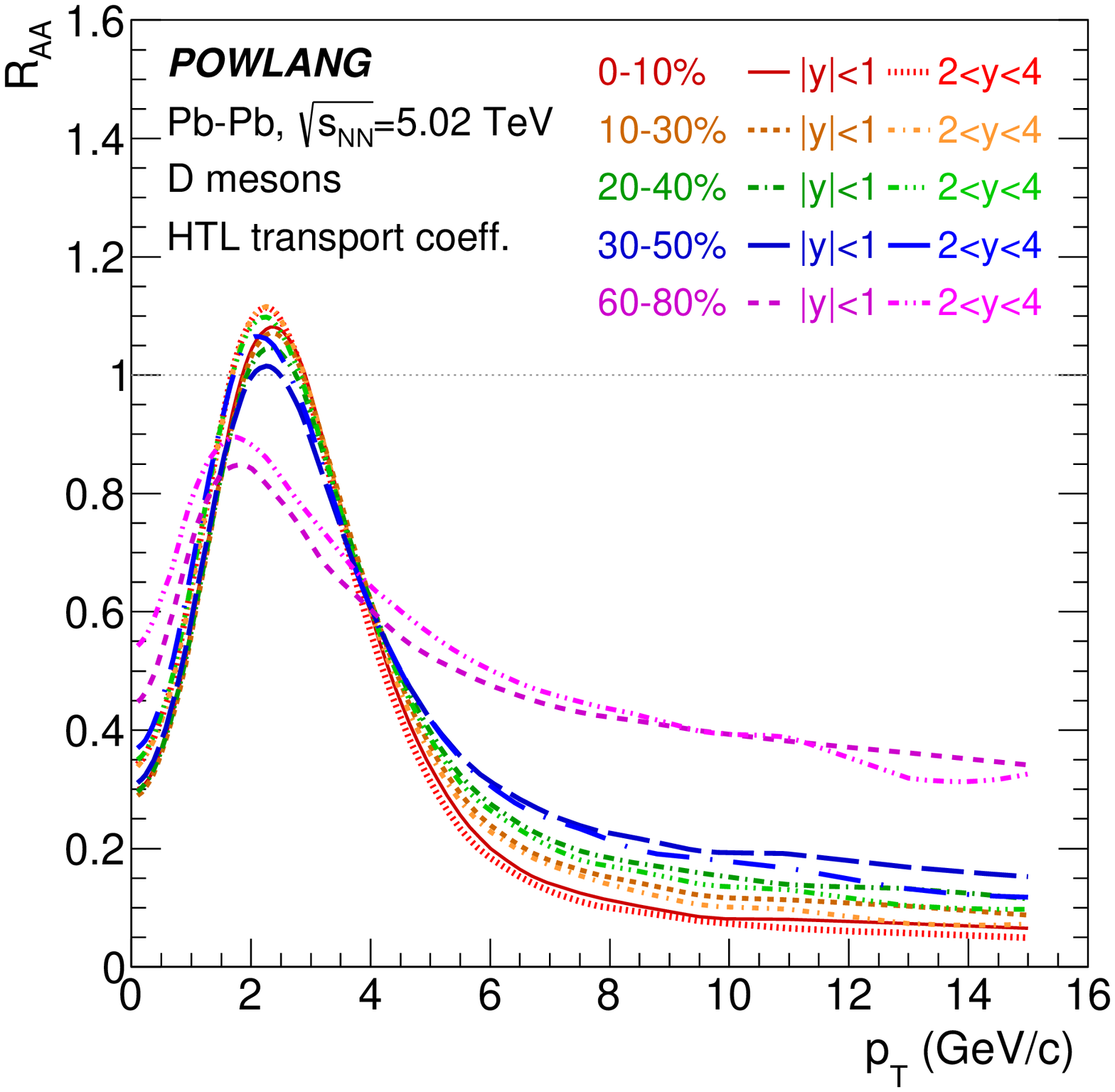}
\includegraphics[clip,width=0.48\textwidth]{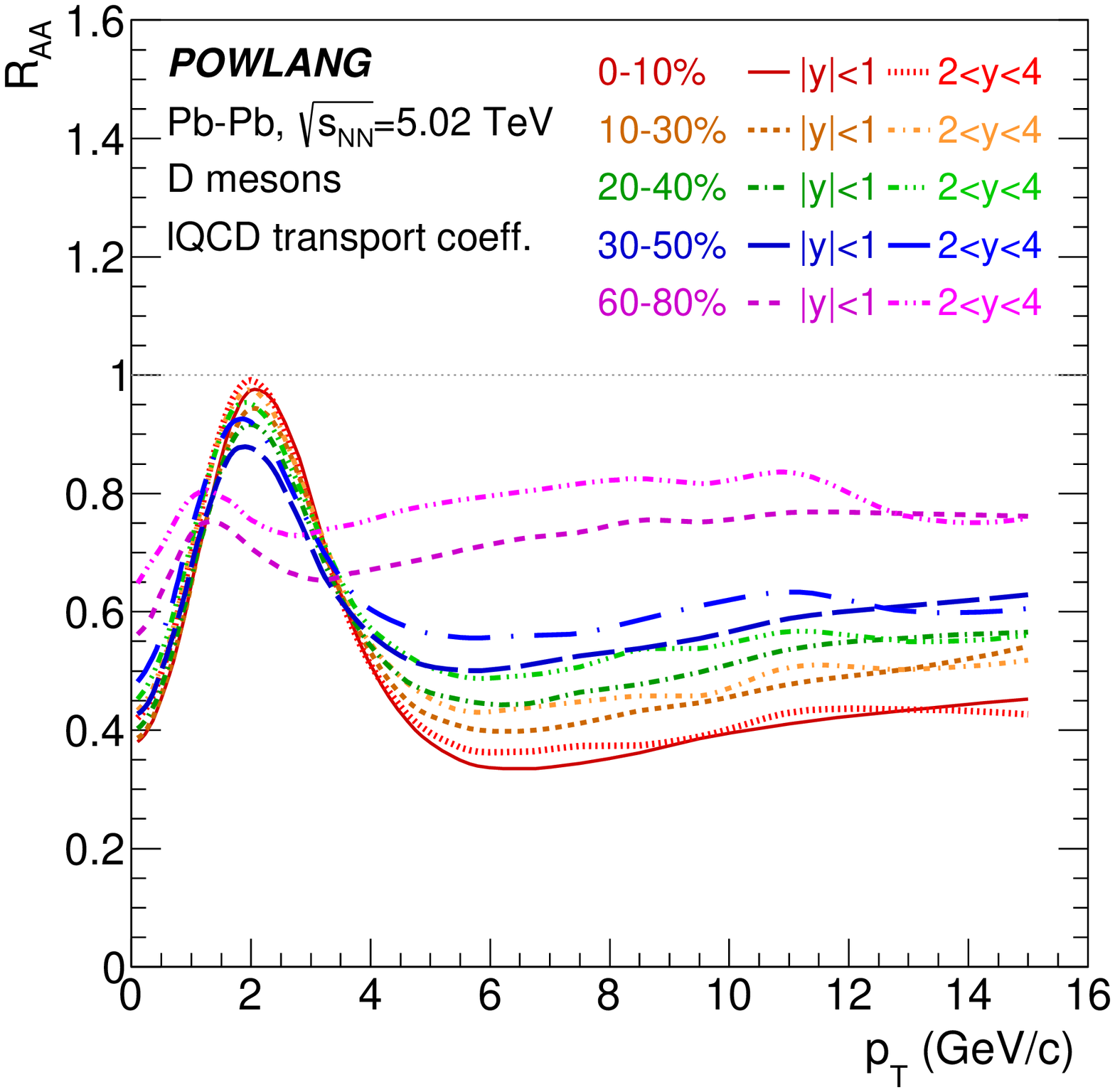}
\caption{The nuclear modification factor $R_{\rm AA}(p_T)$ of charmed hadrons in Pb-Pb collisions for various centrality classes at mid ($|y|<1$) and forward ($2<|y|<4$) rapidity. Both for HTL (left panel) and l-QCD (right panel) transport coefficients POWLANG results display only a mild dependence on the particle rapidity.}\label{fig:D-RAA-HTL-vs-lQCD}
\end{center}
\end{figure}
\begin{figure}
\begin{center}
\includegraphics[clip,width=0.48\textwidth]{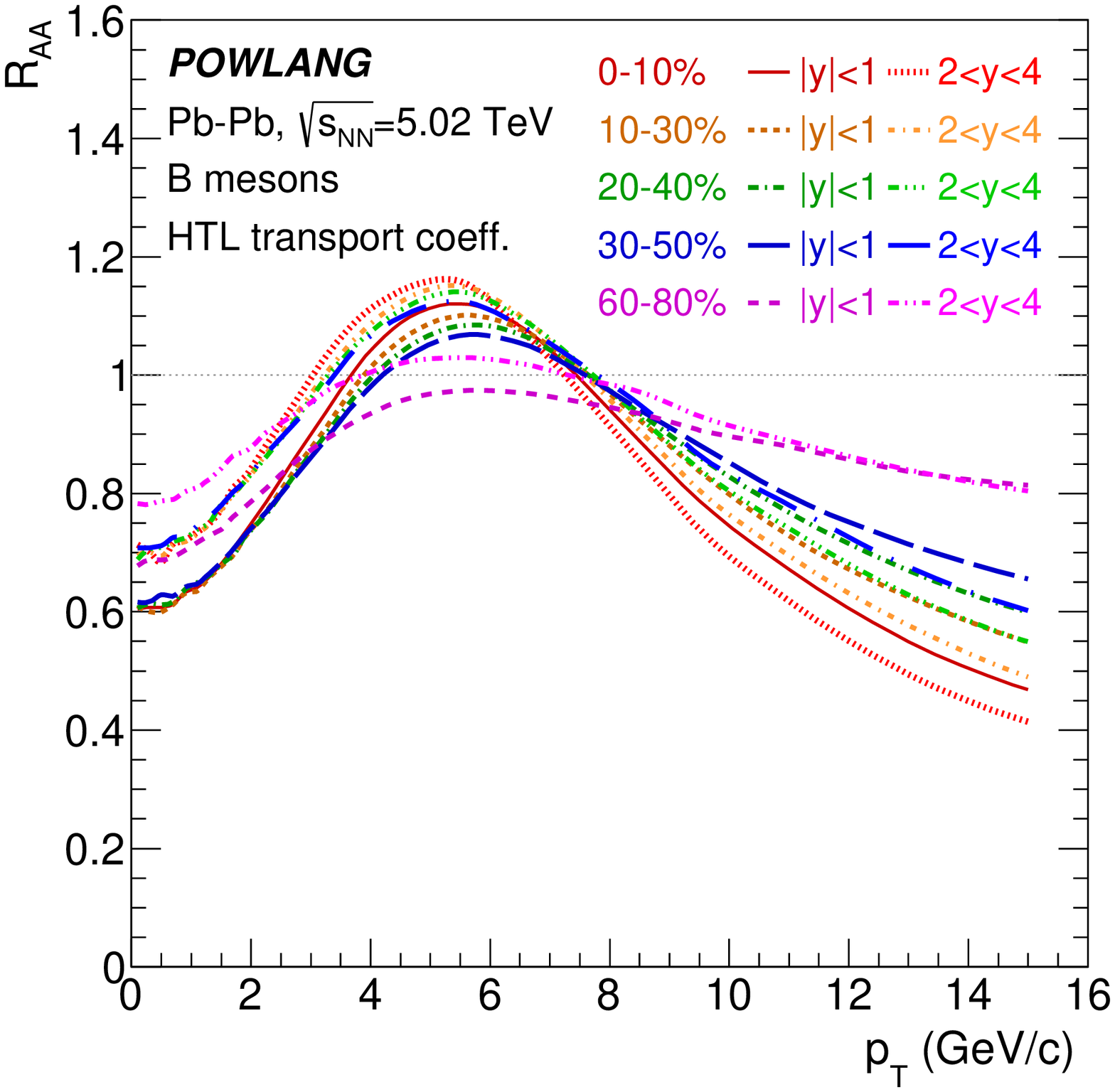}
\includegraphics[clip,width=0.48\textwidth]{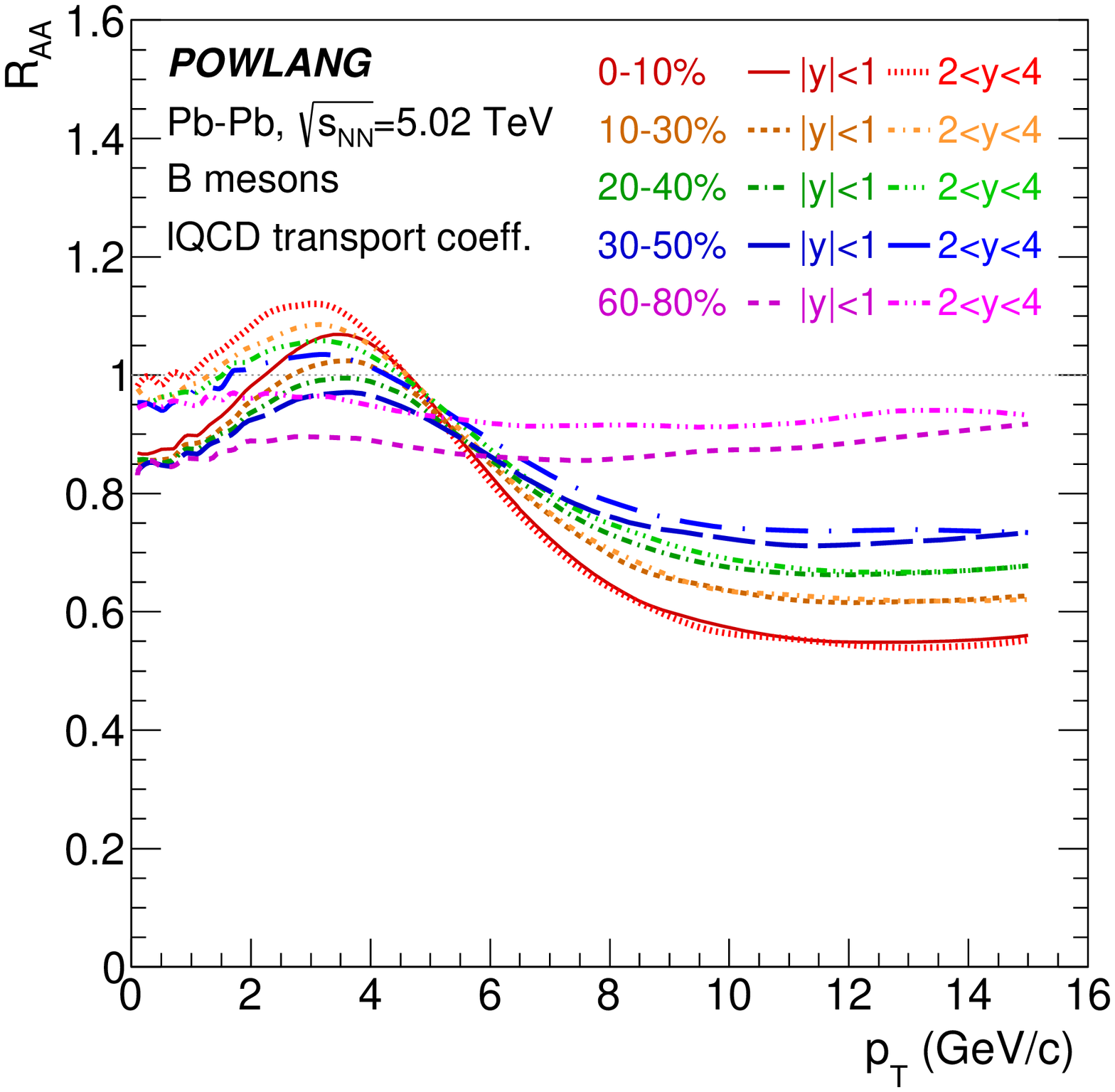}
\caption{The same as in Fig.~\ref{fig:D-RAA-HTL-vs-lQCD} but for $B$-mesons.}\label{fig:B-RAA-HTL-vs-lQCD}
\end{center}
\end{figure}
We start considering the nuclear modification factor of charm and beauty hadrons in Pb-Pb collisions at $\sqrt{s_{\rm NN}}=5.02$ TeV , shown in Fig.~\ref{fig:D-RAA-HTL-vs-lQCD} and~\ref{fig:B-RAA-HTL-vs-lQCD} for different centrality classes. As one can see, results at central $|y|<1$ and forward $2<y<4$ rapidity are very similar. On the contrary, our predictions are quite sensitive to the choice of transport coefficients, in particular for the case of charm. The momentum dependence included in the HTL weak coupling transport coefficients leads to a strong quenching of the spectra at high $p_T$. On the contrary the $R_{\rm AA}$ obtained with l-QCD transport coefficients display a milder quenching of the production of high-momentum HF particles. In both cases the $R_{\rm AA}$ is characterized by an enhancement at moderate $p_T$, reflecting the radial flow of HF hadrons, partially inherited from the recombination with light partons at hadronization.

\begin{figure}[!ht]
\begin{center}
\includegraphics[clip,width=0.48\textwidth]{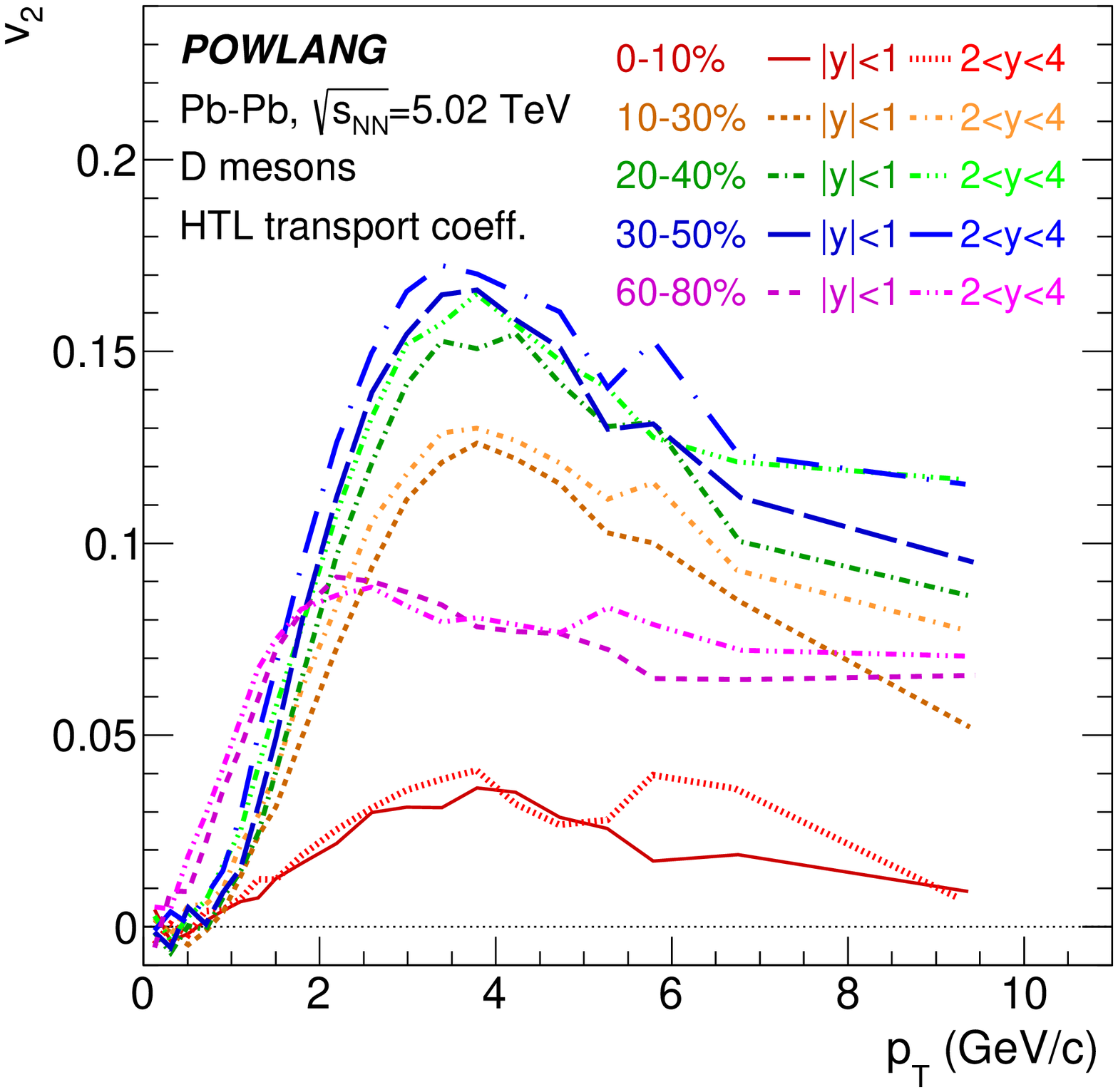}
\includegraphics[clip,width=0.48\textwidth]{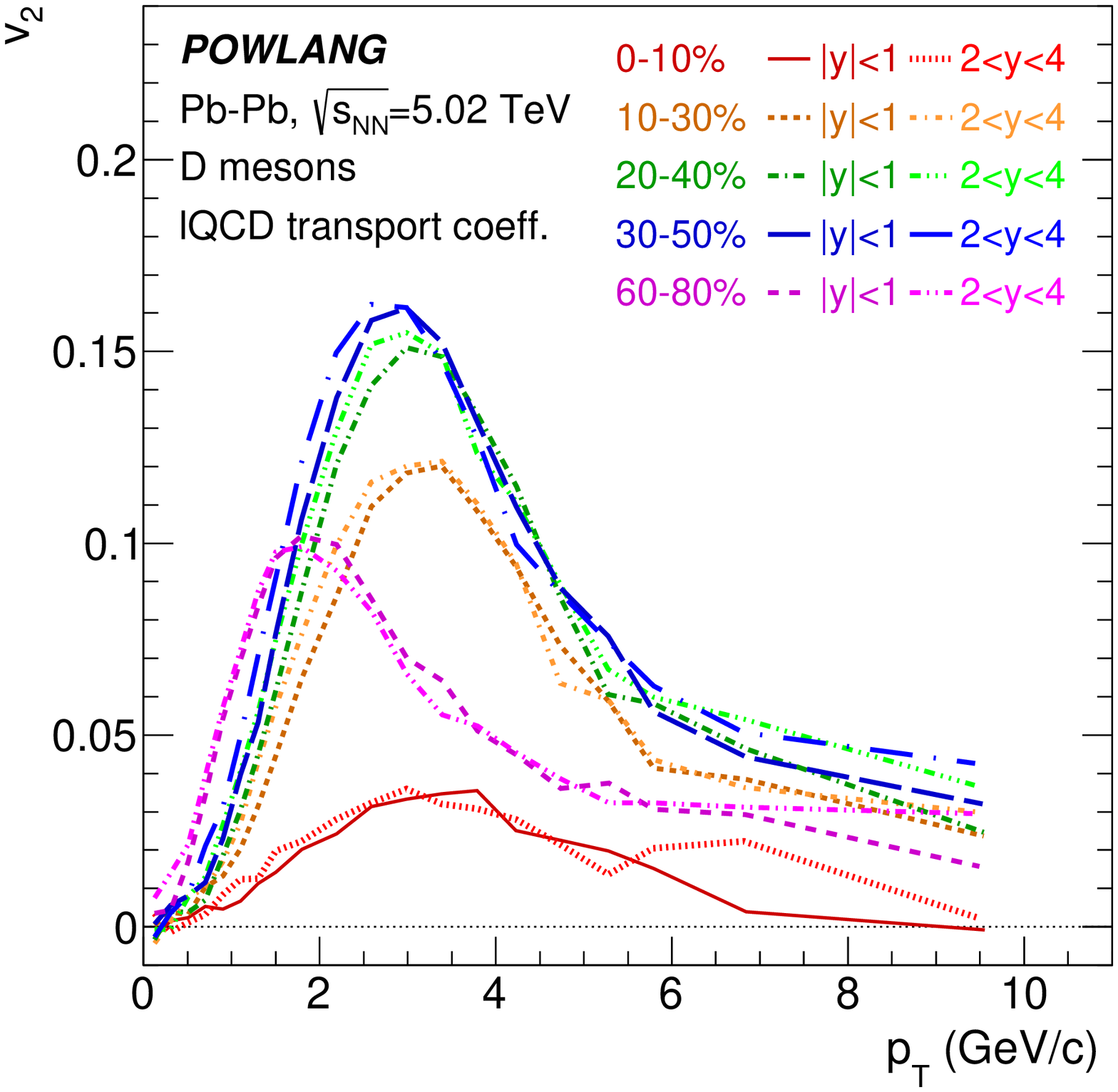}
\caption{The elliptic flow $v_2(p_T)$ of charmed hadrons in Pb-Pb collisions for various centrality classes at mid ($|y|<1$) and forward ($2<|y|<4$) rapidity. Both for HTL (left panel) and l-QCD (right panel) transport coefficients POWLANG results display only a mild dependence on the particle rapidity.}\label{fig:D-v2-HTL-vs-lQCD}
\end{center}
\end{figure}
\begin{figure}[!ht]
\begin{center}
\includegraphics[clip,width=0.48\textwidth]{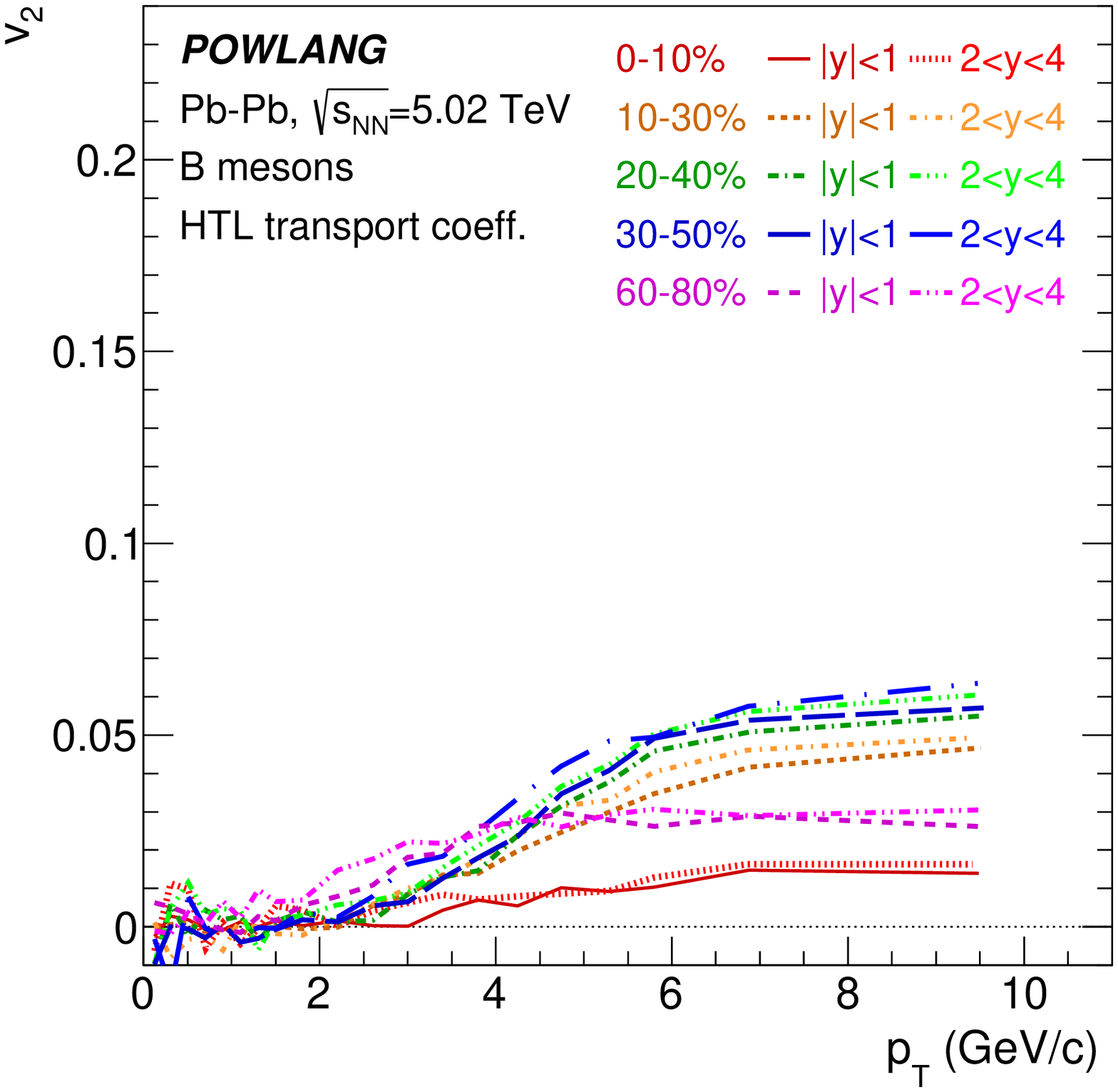}
\includegraphics[clip,width=0.48\textwidth]{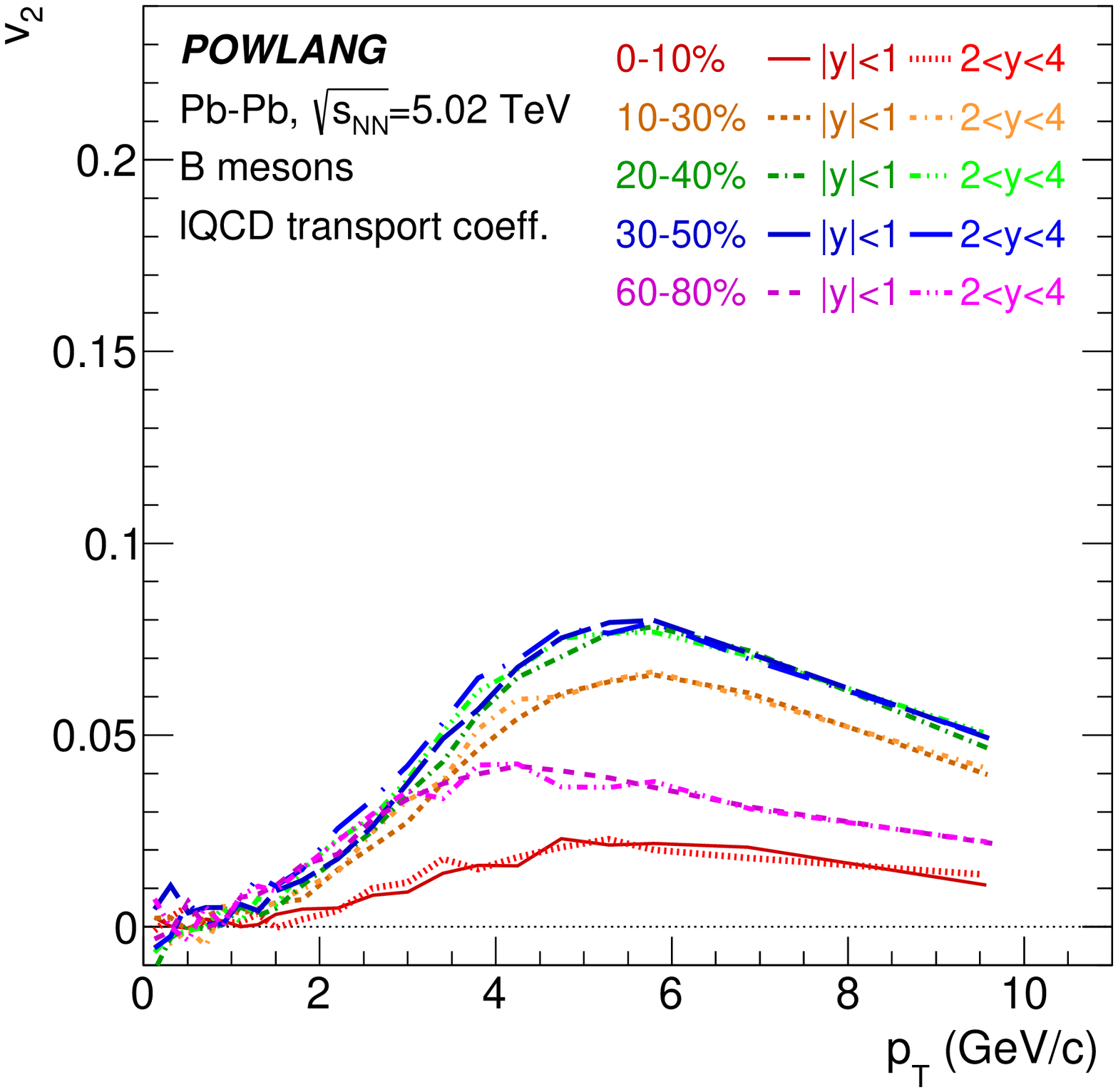}
\caption{The same as in Fig.~\ref{fig:D-v2-HTL-vs-lQCD} but for $B$-mesons.}\label{fig:B-v2-HTL-vs-lQCD}
\end{center}
\end{figure}
We now move to the elliptic flow. Also in this case, both for charm and for beauty, the rapidity dependence of our results is quite mild for all centralities in the considered rapidity intervals $|y|<1$ and $2<|y|<4$. This is consistent with what found for our hydrodynamic background and displayed in the right panel of Fig.~\ref{fig:v2-soft}. We expect that at more forward rapidity the $v_2$ coefficient should actually decrease. 
Concerning charm hadrons, shown in Fig.~\ref{fig:D-v2-HTL-vs-lQCD}, although the magnitude of the $v_2$ is similar both for HTL and l-QCD transport coefficients, in the last case the $v_2$ displays a more rapid decrease with $p_T$. This is consistent with what found for the $R_{\rm AA}$. Actually, at high $p_T$ a positive $v_2$ does not reflect the collective flow of the medium, but simply the different energy-loss suffered by the hard partons moving along or orthogonally to the reaction plane. Hence a milder energy-loss gives rise to a milder azimuthal asymmetry of high-$p_T$ particle production.  
For what concerns beauty, the larger mass of the quarks leads to a smaller azimuthal asymmetry as compared to charm. The elliptic-flow $v_2$ turns out to be larger in the l-QCD case, reflecting the larger value of the momentum diffusion coefficient $\kappa$ entering into the Langevin equation. Notice that in the limited kinematic region considered in the figure, the rise of the HTL transport coefficients for beauty as a function of the HQ momentum is mild and not sufficient to compensate their lower value at rest with respect to the non-perturbative l-QCD estimate.

\begin{figure}[!ht]
\begin{center}
\includegraphics[clip,width=0.98\textwidth]{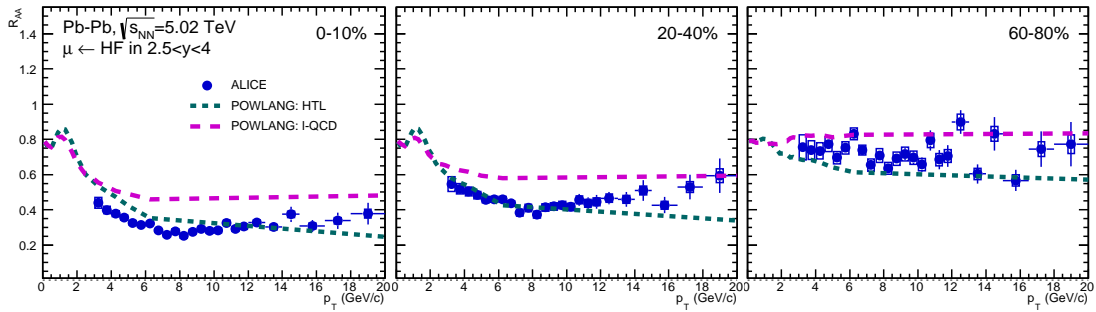}
\caption{The nuclear modification factor of muons from charm and beauty decay in Pb-Pb collisions at forward rapidity provided by our simulations with HTL (short-dashed) and l-QCD (long-dashed) transport coefficients. Numerical findings are compared to ALICE data~\cite{Abelev:2012qh,Acharya:2020xzh}}.
\label{fig:RAA_muDB}
\end{center}
\end{figure}
\begin{figure}[!ht]
\begin{center}
\includegraphics[clip,width=0.98\textwidth]{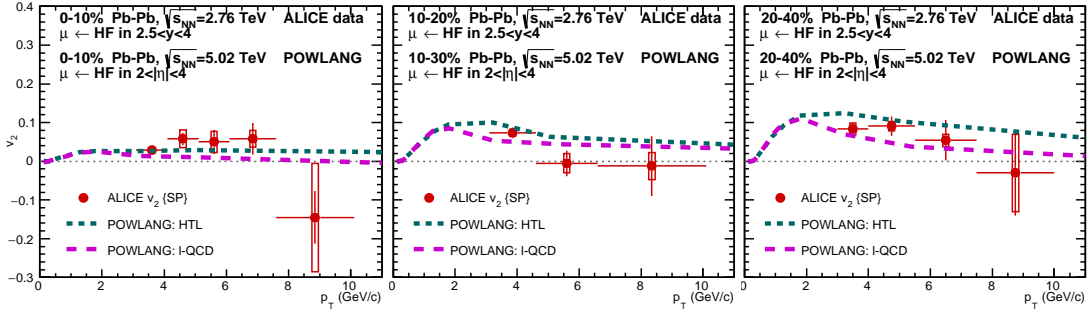}
\caption{The same as in Fig.~\ref{fig:RAA_muDB} but for the elliptic-flow coefficient $v_2$. Results are compared to the so far available ALICE data at forward rapidity at lower center-of-mass energy~\cite{Adam:2015pga}.}\label{fig:v2_muDB}
\end{center}
\end{figure}
We finally show our predictions for the muons arising from the semileptonic decays of charm and beauty hadrons, so far the only available observable to get access to open-HF production at forward rapidity in nucleus-nucleus collisions.
In Figs.~\ref{fig:RAA_muDB} and~\ref{fig:v2_muDB} we show our inclusive ($\mu\leftarrow c\,+\,\mu\leftarrow b$) results for their nuclear modification factor and elliptic flow in Pb-Pb collisions at $\sqrt{s_{\rm NN}}=5.02$ TeV at various centralities. As usual, the results obtained with HTL and l-QCD transport coefficients are displayed and compared to experimental data obtained by the ALICE collaboration~\cite{Abelev:2012qh,Adam:2015pga,Acharya:2020xzh}. Medium effects look milder in the l-QCD case, but this has to be interpreted as arising from our lack of information from non-perturbative calculations on the momentum dependence of the HQ transport coefficients, which is then neglected. This is particularly relevant for high-$p_T$ leptons, since they arise from the decays of parent HF hadrons of even larger momentum. 

\begin{figure}[!ht]
\begin{center}
\includegraphics[clip,width=0.98\textwidth]{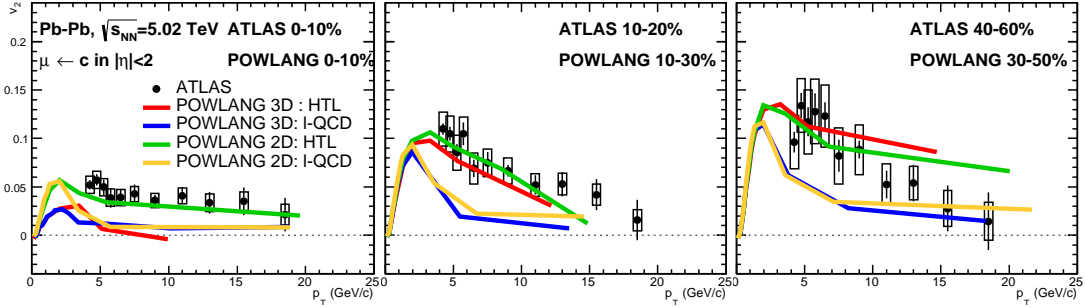}
\caption{The elliptic flow coefficient of muons from charm decays in Pb-Pb collisions provided by our simulations with HTL and l-QCD transport coefficients, starting from 3D optical-Glauber and 2D MC-Glauber initial conditions. Results at central rapidity are compared to ATLAS data~\cite{Aad:2020grf}.}\label{fig:v2_muD}
\end{center}
\end{figure}
\begin{figure}[!ht]
\begin{center}
\includegraphics[clip,width=0.98\textwidth]{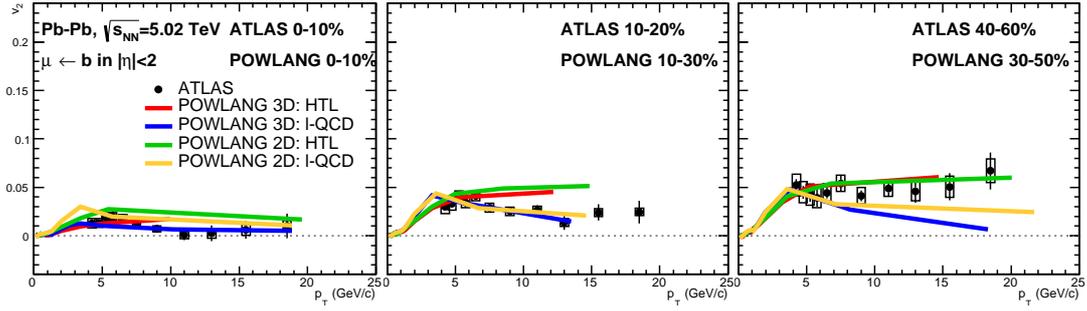}
\caption{The same as in Fig.~\ref{fig:v2_muD} but for muons from beauty decays.}\label{fig:v2_muB}
\end{center}
\end{figure}
\begin{figure}[!ht]
\begin{center}
\includegraphics[clip,width=0.98\textwidth]{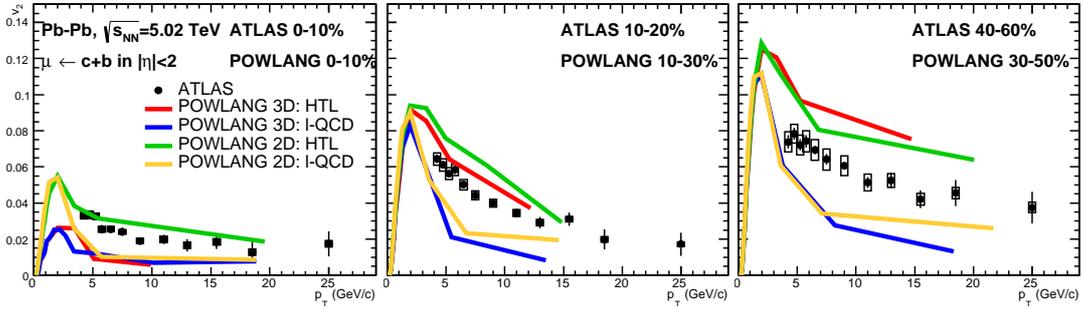}
\caption{The combined results for the $v_2$ coefficient of muons from charm and beauty decays compared to ATLAS data.}\label{fig:v2_muDB_central}
\end{center}
\end{figure}
So far the ALICE experiment does not have the possibity of distinguishing the muons from charm and beauty decays. Such a more differential analysis for the azimuthal asymmetry of HF-muon production in Pb-Pb collisions was instead performed by the ATLAS collaboration~\cite{Aad:2020grf}, although limited to central rapidity, $|y|<2$.
Hence, in Figs.~\ref{fig:v2_muD} and~\ref{fig:v2_muB} we display the separate findings for the elliptic-flow coefficient $v_2$ of muons from charm and beauty decays obtained with our transport calculations. Their combined results are shown in Fig.~\ref{fig:v2_muDB_central}. 
Since ATLAS measurements are carried out in a quite broad range around mid-rapidity, we also show the predictions of (2+1)D transport simulations, assuming longitudinal boost-invariance for the hydrodynamic background and starting from the Monte-Carlo Glauber initial conditions discussed in Ref.~\cite{Beraudo:2017gxw,Beraudo:2018tpr}. As expected, the event-by-event fluctuations accounted for by the MC-Glauber initialization make the initial eccentricity of central collisions larger than the one provided by the optical Glauber model. {Furthermore in the MC-Glauber initialization employed in Refs.~\cite{Beraudo:2017gxw,Beraudo:2018tpr} the initial entropy deposition in the transverse plane is assumed to arise entirely from the binary nucleon-nucleon collisions (i.e. $\alpha_h\!=\!1$) and this also contributes to make the initial eccentricity larger. The comparison of the results for the average elliptic deformation of the fireball $\epsilon_2\equiv\langle y^2-x^2\rangle/\langle y^2+x^2\rangle$ at the beginning of its evolution provided by the two different initializations is shown in Table~\ref{table:eps2}. In the MC-Glauber case $\epsilon_2$ is measured in a rotated frame in which the event-plane angle $\Psi_2$ lies along the $x$-axis. The effect of the event-by-event fluctuations is particularly strong in the 0-10\% centrality class.
Accordingly, for a given set of transport coefficients, the elliptic flow of HF decay muons in central ($0-10\%$) Pb-Pb collisions obtained with our (2+1)D transport simulations is larger than what found with the (3+1)D calculations, so far limited to an initial geometry given by the optical Glauber model.
Notice that ATLAS data for HF decay muons extend to much larger transverse momenta than ALICE results. As already discussed, for such large values of $p_T$ of the muons -- which arise from the decays of parent HF hadrons of even higher momentum -- the non-vanishing $v_2$ does not reflect the collective flow of the background fireball but rather the different amount of energy loss suffered by HQ's of very high energy propagating along or orthogonally to the reaction plane. In this kinematic range -- in which however also other processes like medium-induced gluon radiation should be considered -- the very different behaviour of the HTL and l-QCD transport coefficients explains the larger $v_2$ obtained in the HTL case.} 
\begin{table}[!h]
\begin{center}
\begin{tabular}{|c|c|c|}
\hline
{} & MC-Glauber, $\alpha_h\!=\!1$~\cite{Beraudo:2017gxw,Beraudo:2018tpr} & Opt-Glauber, $\alpha_h\!=\!0.15$ (this work)\\
\hline
0-10\% & $\langle\epsilon_2\rangle=0.142$ & $\langle\epsilon_2\rangle=0.063$\\
\hline
10-30\% & $\langle\epsilon_2\rangle=0.286$ & $\langle\epsilon_2\rangle=0.223$\\
\hline
30-50\% & $\langle\epsilon_2\rangle=0.438$ & $\langle\epsilon_2\rangle=0.350$\\
\hline
\end{tabular}
\caption{The initial average geometric eccentricity of the system produced in Pb-Pb collisions at $\sqrt{s_{\rm NN}}=5.02$ TeV for different kinds of initialization and centrality classes. In the MC-Glauber case, in each event, the elliptic deformation $\epsilon_2$ of the fireball is measured with respect to the event-plane angle $\Psi_2$, which does not necessarily lie in the reaction plane.}\label{table:eps2}
\end{center}
\end{table}


\section{Discussion and prospects}\label{sec:discussion}
A full (3+1)D realistic modelling of the medium is necessary to describe several observables in relativistic heavy-ion collisions, whose dependence on rapidity has at the same time the potential to get access to a richer information on the initial state of the system and to provide additional constraints to the transport coefficients of the medium. First of all, the produced fireball has a finite longitudinal extension, with an upper bound set by the rapidity of the colliding nuclei, but with a non-negligble energy density deposited only in a more limited range in space-time rapidity around $\eta_s=0$; hence, medium effects on the final particle distributions can be different in the forward and central regions.
Secondly, the fact that participant nucleons tend to deposit a fraction of their energy mostly along their direction of motion rather then backwards leads to a tilting of the initial geometry of the medium in the $\eta_s-x$ plane. 
In this connection the asymmetry of the pressure gradients in the $\eta_s-x$ plane leads to the development of a non-vanishing directed flow $v_1$ of light hadrons, which is however quite small.

The non-trivial longitudinal shape of the initial conditions affects also HF observables, which are the subject of this work, in which transport equations are interfaced to the output of full (3+1)D hydrodynamic calculations.
The different production mechanism of HF particles with respect to the one of the light thermal partons of the fireball has the potential to provide further insight into several features of the medium, in particular concerning its initial stage, due to the very short formation time of HQ's.   
Our major finding is that also $D$ mesons develop a non-vanishing directed-flow coefficient $v_1$, which turns out to be at least one order of magnitude larger than the one of light hadrons. 
This occurrence was found also in previous independent studies~\cite{Chatterjee:2017ahy,Das:2016cwd} and cannot be interpreted as simply due to the collective flow acquired through the interaction with the medium. Its origin rather lies in the mismatch between the initial spatial distribution of heavy quarks and the energy density of the medium, the latter characterized by the above mentioned tilting in the $\eta_s-x$ plane. Remarkably, the stronger the coupling of the heavy quarks with the medium the smaller their spatial diffusion coefficient $D_s$ is, hence such an asymmetry tends to persist throughout the evolution of the medium, since in this case each heavy quark tends to move following its original fluid cell. As a consequence, while in the limit of very strong coupling the momentum distribution and the elliptic flow of the heavy quarks tend to the Cooper-Frye result expected for particles at local thermal equilibrium, the directed-flow coefficient $v_1$ displays a completely different behaviour, tending to a much larger value. This is due to the fact that, although each heavy quark flows with the fluid, as a consequence of the initial mismatch between the HF and bulk-medium spatial distribution, the different cells belonging to the decoupling hypersurface are characterized by an over-population or under-population of charm quarks with respect to the case of full thermodynamic equilibrium. 

Although the major motivation of our work was the study of the $D$-meson directed flow, we also applied our setup to model medium effects on the HF particle distributions at forward rapidity, finding, overall, only small differences with respect to the results obtained in the central region.
Concerning the comparison with the experimental data, the measurements available to date are based on the detection of forward muons from charm and beautys decay performed by the ALICE experiment. The current impossibility of disentangling the two contributions will be overcome through the upgrade of the experimental apparatus. Also future measurements by the LHCb collaboration at forward rapidity will provide important information in this kinematic domain. 

A possible relevant issue to investigate in the near future is the effect of the initial strong magnetic fields present in non-central heavy-ion collisions on the final distributions of heavy-quarks, produced in the very early stage in hard scattering processes. Past studies suggest~\cite{Das:2016cwd,Chatterjee:2018lsx} that one should look for differences in the $v_1$ coefficient between $D^0$ and $\overline{D^0}$ mesons, which can only arise from the partonic stage of the fireball evolution. The experimental situation is not yet completely settled~\cite{Adam:2019wnk,Acharya:2019ijj}, with some hints of a possible splitting found by the ALICE collaboration, but with the need of reducing the current experimental uncertainties and of exploring the effect in different centrality classes.
Having shown that, with our transport model, we are able to account for the rapidity dependence of several HF observables and, in particular, to reproduce the size of the average $D$-meson directed-flow coefficient observed by the STAR collaboration, as a next step of our study we plan to insert also the electromagnetic fields into our transport equations, coupling them to a full RMHD code that some of us contributed to develop in the past~\cite{Inghirami:2016iru,Inghirami:2019mkc} and which provides a consistent solution of the in-medium Maxwell equations. This will allow one to get access to the initial magnetic field and, extending the study to the resistive MHD case, to the electric conductivity of the QGP. We leave the above interesting topic for a forthcoming independent publication. 
  
\appendix
\section{Validation of the 3+1 transport simulations}
\begin{figure}[!ht]
\begin{center}
\includegraphics[clip,width=0.48\textwidth]{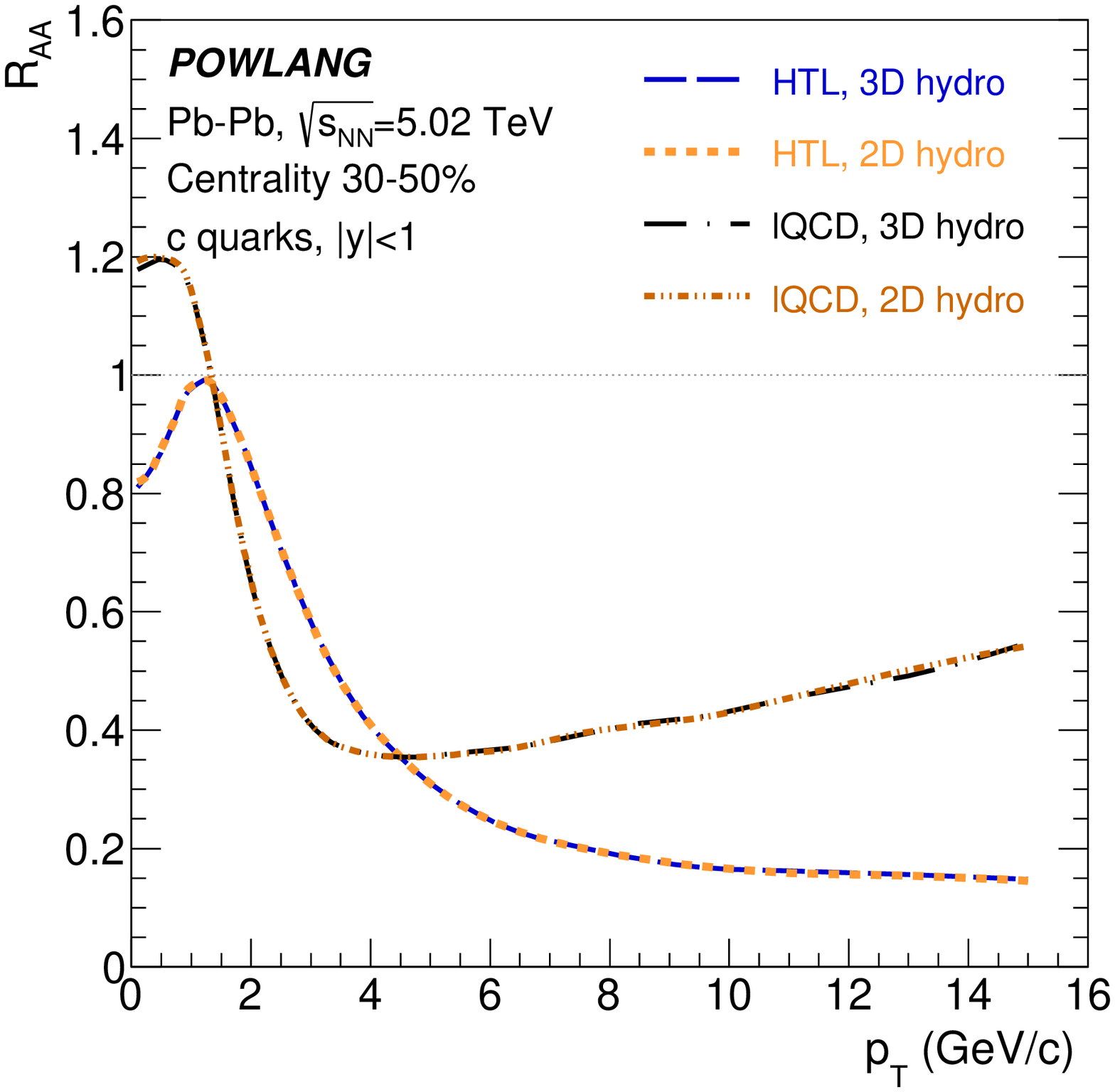}
\includegraphics[clip,width=0.48\textwidth]{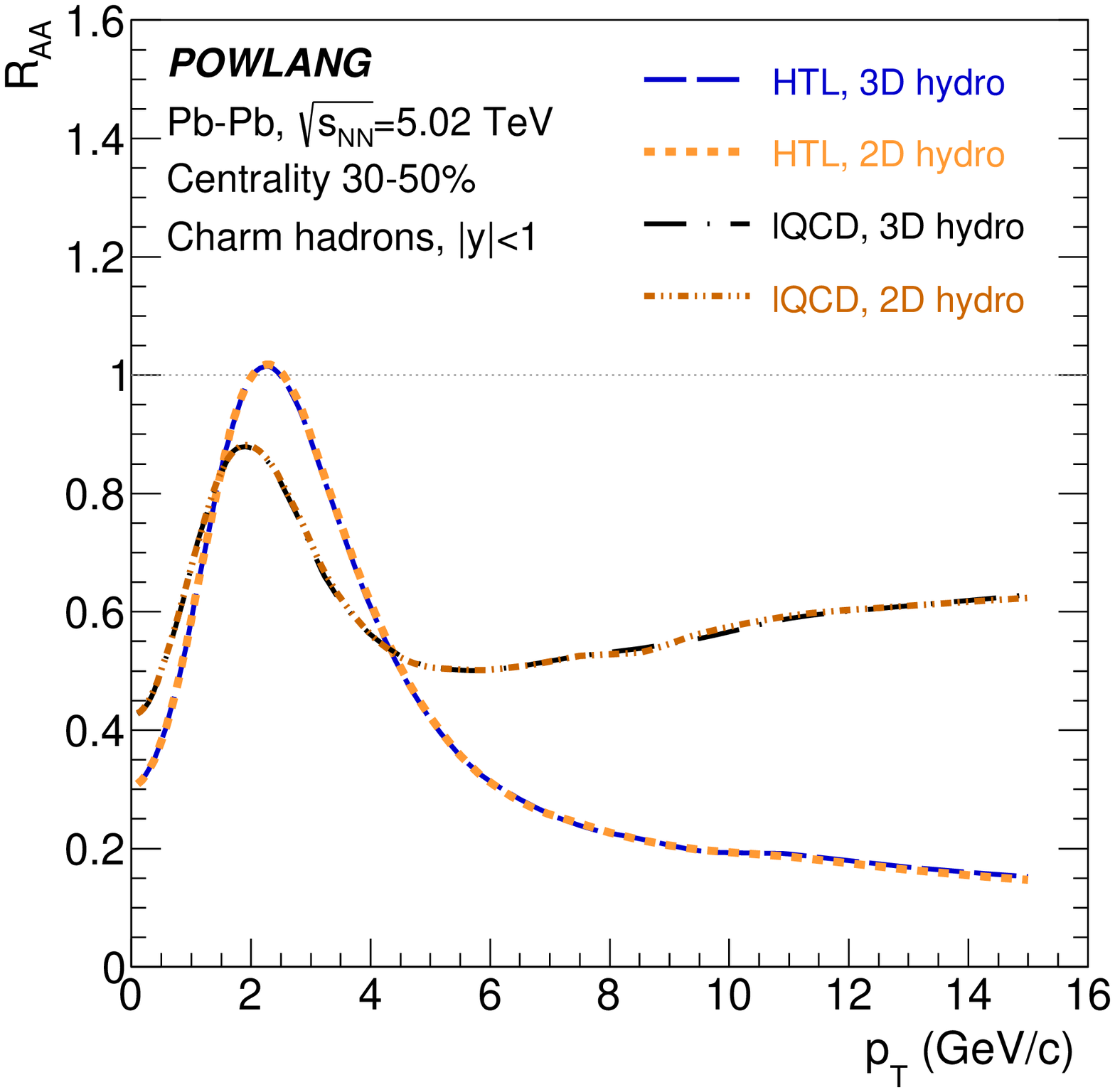}
\caption{Comparison of the nuclear modification factor of charm quarks (left panel) and hadrons (right panel) at mid-rapidity in 2D and 3D transport simulation.}\label{fig:validation-RAA}
\end{center}
\end{figure}
\begin{figure}[!ht]
\begin{center}
\includegraphics[clip,width=0.48\textwidth]{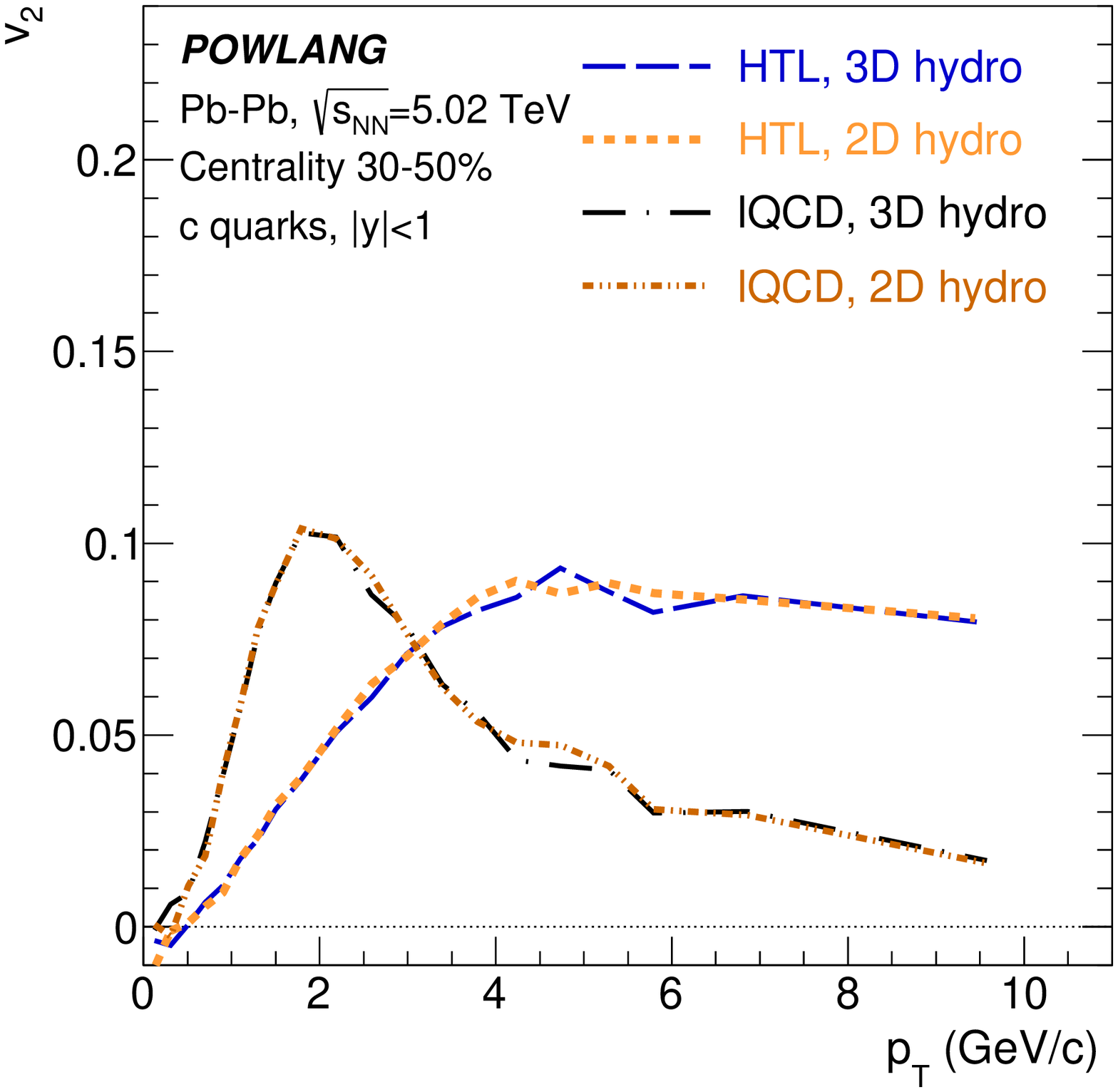}
\includegraphics[clip,width=0.48\textwidth]{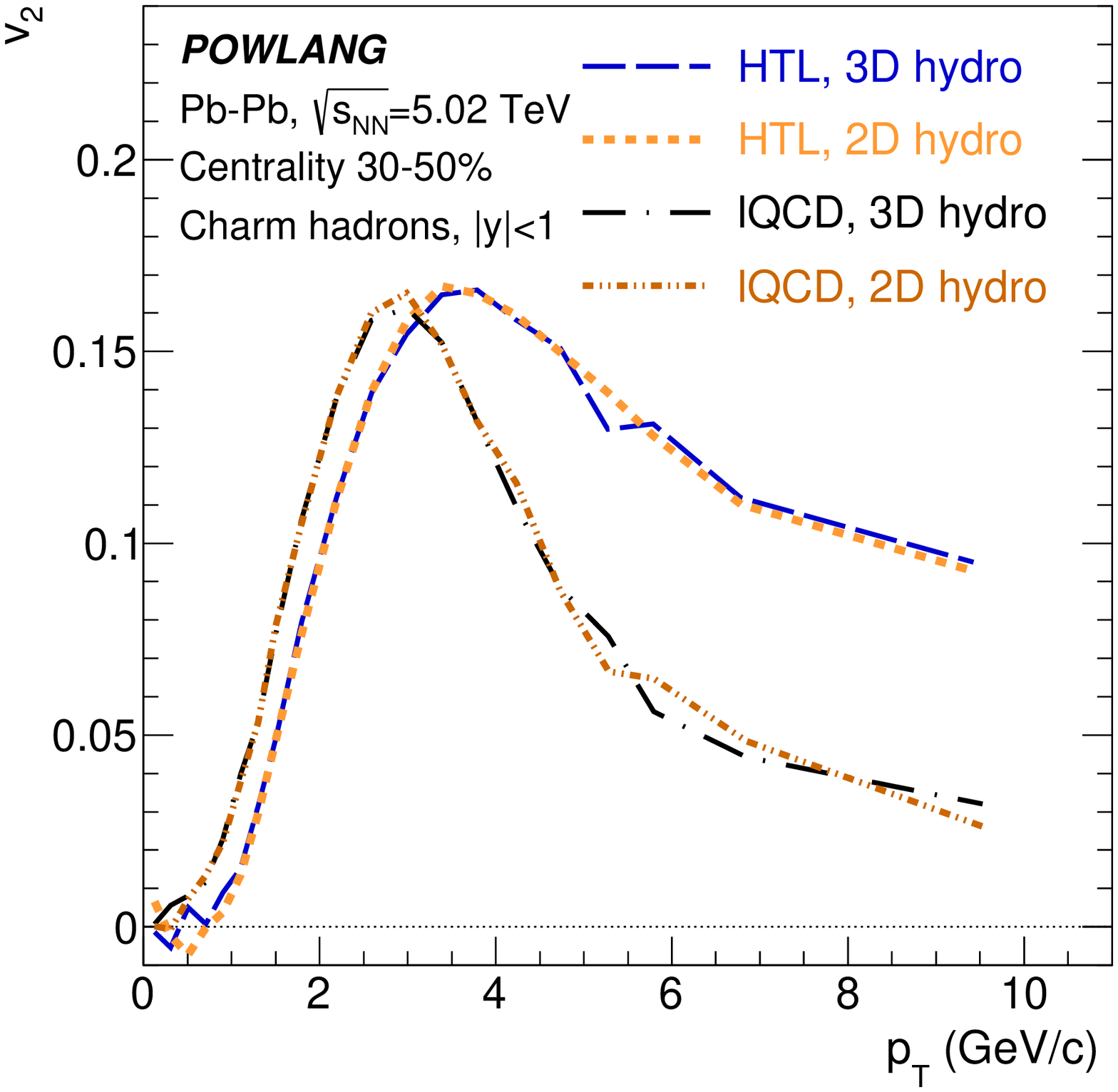}
\caption{Comparison of the elliptic flow coefficient $v_2$ of charm quarks (left panel) and hadrons (right panel) at mid-rapidity in 2D and 3D transport simulation.}\label{fig:validation-v2}
\end{center}
\end{figure}
Full 3+1 transport simulations require storing a huge amount of information. Hence we performed our hydrodynamic calculations employing a grid of $101\times 101\times 101$ cells in the $x,\,y$ and $\eta_s$ directions, ranging from -16.1 to 16.1 fm in the transverse plane and from -12.1 to 12.1 in space-time rapidity. A time-step $\Delta\tau\!=\!0.1$ fm/c in longitudinal proper time was used to store the information on the temperature and velocity of the fluid. This corresponds to a broader grid with respect to the $161\times 161\times 1$ one, with time-step $\Delta\tau=0.04$ fm/c, usually employed in our past 2D simulations. Here we verify that this does not affect our findings for observables around mid-rapidity when results of 2D and 3D simulations with the same initial conditions in the transverse plane are compared. This is clearly shown in Figs.~\ref{fig:validation-RAA} and~\ref{fig:validation-v2} for the nuclear modification factor and elliptic flow of charm in non-central Pb-Pb collisions, respectively. The figures show also that, as long as observables integrated over a not too large symmetric interval around mid-rapidity are concerned, the non-vanishing longitudinal pressure gradients of the fireball in the 3+1 case do not significantly affect the results.

\bibliography{draft}

\end{document}